\documentclass[aps,prx,twocolumn,nopacs,superscriptaddress]{revtex4-1}

\usepackage[breaklinks=true,colorlinks,citecolor=blue,linkcolor=blue,urlcolor=blue]{hyperref}
\usepackage{epsfig,mathrsfs,color,latexsym,subfigure,marginnote,graphicx,verbatim,relsize,mathrsfs,color,array,amsmath,amsfonts,amssymb,graphicx}
\usepackage{xspace}
\newcommand{\rrscan}{r\textsuperscript{2}SCAN\xspace}

\newcommand{\SB}[1] {{\it{\color{blue}#1}}}
\newcommand{\veck}{\boldsymbol{k}}
\newcommand\vecq{\boldsymbol{q}}

\begin{document}

\title{Accurate Electron-phonon Interactions from Advanced Density Functional Theory}

\author{Yanyong~Wang}
\thanks{These authors contributed equally to this work}
\affiliation{Department of Physics and Engineering Physics, Tulane University, New Orleans, LA 70118, USA}

\author{Manuel~Engel}
\thanks{These authors contributed equally to this work}
\affiliation{VASP Software GmbH, Berggasse 21/14, 1090 Vienna, Austria}

\author{Christopher Lane}
\affiliation{Theoretical Division, Los Alamos National Laboratory, Los Alamos, New Mexico 87545, USA}

\author{Henrique Miranda}
\affiliation{VASP Software GmbH, Berggasse 21/14, 1090 Vienna, Austria}

\author{Lin Hou}
\affiliation{Department of Physics and Engineering Physics, Tulane University, New Orleans, LA 70118, USA}
\affiliation{Theoretical Division, Los Alamos National Laboratory, Los Alamos, New Mexico 87545, USA}

\author{Bernardo~Barbiellini}
\affiliation{Department of Physics, School of Engineering Science, LUT University, FI-53850 Lappeenranta, Finland}
\affiliation{Department of Physics, Northeastern University, Boston, MA 02115, USA}
\affiliation{Quantum Materials and Sensing Institute, Northeastern University, Burlington, MA 01803, USA}

\author{Robert S. Markiewicz}
\affiliation{Department of Physics, Northeastern University, Boston, MA 02115, USA}
\affiliation{Quantum Materials and Sensing Institute, Northeastern University, Burlington, MA 01803, USA}

\author{Jian-Xin Zhu}
\affiliation{Theoretical Division, Los Alamos National Laboratory, Los Alamos, New Mexico 87545, USA}

\author{Georg Kresse}
\affiliation{VASP Software GmbH, Berggasse 21/14, 1090 Vienna, Austria}
\affiliation{University of Vienna, Faculty of Physics, Kolingasse 14-16, A-1090 Vienna, Austria}

\author{Arun~Bansil}
\email[Corresponding author:~]{ar.bansil@neu.edu}
\affiliation{Department of Physics, Northeastern University, Boston, MA 02115, USA}
\affiliation{Quantum Materials and Sensing Institute, Northeastern University, Burlington, MA 01803, USA}

\author{Jianwei Sun}
\email[Corresponding author:~]{jsun@tulane.edu}
\affiliation{Department of Physics and Engineering Physics, Tulane University, New Orleans, LA 70118, USA}

\author{Ruiqi~Zhang}
\thanks{These authors contributed equally to this work}
\email[Corresponding author:~]{rzhang16@tulane.edu}
\affiliation{Department of Physics and Engineering Physics, Tulane University, New Orleans, LA 70118, USA}

\begin{abstract}

Electron-phonon coupling (EPC) is key for understanding many properties of materials such as superconductivity and electric resistivity. Although first principles density-functional-theory (DFT) based EPC calculations are used widely, their efficacy is limited by the accuracy and efficiency of the underlying exchange-correlation functionals. These limitations become exacerbated in complex $d$- and $f$-electron materials, where beyond-DFT approaches and empirical corrections, such as the Hubbard $U$, are commonly invoked. Here, using the examples of CoO and NiO, we show how the efficient \rrscan density functional correctly captures strong EPC effects in transition-metal oxides without requiring the introduction of empirical parameters.  We also demonstrate the ability of \rrscan to accurately model phonon-mediated superconducting properties of the main group compounds (e.g., MgB$_2$), with improved electronic bands and phonon dispersions over those of traditional density functionals. Our study provides a pathway for extending the scope of accurate first principles modeling of electron-phonon interactions to encompass complex $d$-electron materials.

\end{abstract} 

\maketitle

\SB{Introduction}\textbf{\---}Electron-phonon coupling (EPC) is fundamental to understanding a range of phenomena in condensed matter physics and materials science~\cite{FelicianoRMP2017,grimvall1981electron}, including phonon-mediated superconductivity~\cite{Pellegrini2024NRP}, electrical resistivity in metals, carrier lifetime and mobility in semiconductors, and the temperature-dependent behavior of electronic structures~\cite{ManuelPRB2020,ManuelPRB2022}. The development of Kohn-Sham (KS) density functional theory (DFT) has enabled the quantitative prediction of EPCs through first-principles calculations, which have advanced significantly over the last few decades~\cite{FelicianoRMP2017}. The two main methods that exist for calculating EPC matrix elements are density functional perturbation theory (DFPT)~\cite{PhononDFPTRMP2001} and the finite-difference method~\cite{ShanghJCPeph,Monserrat2018FD}.

DFPT, often combined with maximally localized Wannier functions~\cite{EPW2007,Perturbo}, provides an efficient means of calculating EPC matrix elements across the entire Brillouin Zone (BZ). However, it is typically limited to traditional density functional approximations (DFAs) that are not explicitly dependent on KS orbitals, such as the local-density approximation (LDA) and generalized-gradient approximations (GGAs). The finite-difference approach offers greater flexibility, allowing the inclusion of orbital-dependent density functionals and integration with advanced electronic structure solvers that consider correlation effects beyond the mean-field description~\cite{Monserrat2018FD,FDGWdiamond,ZhipingPRX2013}, but it often requires large supercells, making it computationally intensive.

For complex materials with \( d \)- and \( f \)-electrons, LDA and GGAs often fail to accurately capture their electronic structures~\cite{Furness2018,Lane2018,Zhang2020b,ZhangPRLLaNiO2} and phonon properties~\cite{NingJCPDFAYBCO6,NingCMPhonon,NingPRBYBCO6,DFPTUeph}, resulting in inaccurate predictions for EPCs. While methods like DFPT+$U$ can improve accuracy~\cite{DFPTUeph,DFPTUCoOCeltech}, they rely on parameterizations specific for $d$- and $f$-electrons. More precise approaches, such as combining GW with DFPT~\cite{ZhengLuGWPTBaTiO3,ZhengluGWPTCuprate}, further enhance EPC predictions but involve substantial computational costs. This underscores the need for efficient methods that can deliver accurate EPC predictions without the computational burden of beyond-DFT methods.

Meta-GGAs like the strongly-constrained and appropriately-normed (SCAN) ~\cite{Sun2015} density functional and its variant \rrscan~\cite{R2SCANJPCL}, which are orbital dependent and scale computationally similar as LDA and GGAs, have shown promise in providing accurate descriptions of total energies and electronic structures across diverse systems~\cite{Sun2015,R2SCANJPCL,sun2016accurate}. These functionals have proven effective in materials ranging from liquid water~\cite{chen2017ab} to metal oxides~\cite{gautam2018evaluating}, and complex materials with $d$- and $f$-electrons~\cite{Zhang2020b,ZhangR2020}, like the cuprates~\cite{Lane2018, Furness2018, Zhang2020} and nickelates~\cite{Zhang2021,Lane2023,ZhangPRLLaNiO2}. Recent work also indicates that SCAN/\rrscan can accurately calculate phonon dispersions of cuprates~\cite{Zhang2020b,NingPRBYBCO6} and NiO~\cite{NingCMPhonon}. However, their performance on EPC predictions remains unexplored.

\begin{figure*}[htpb]
\centering
\includegraphics[width=0.99\linewidth]{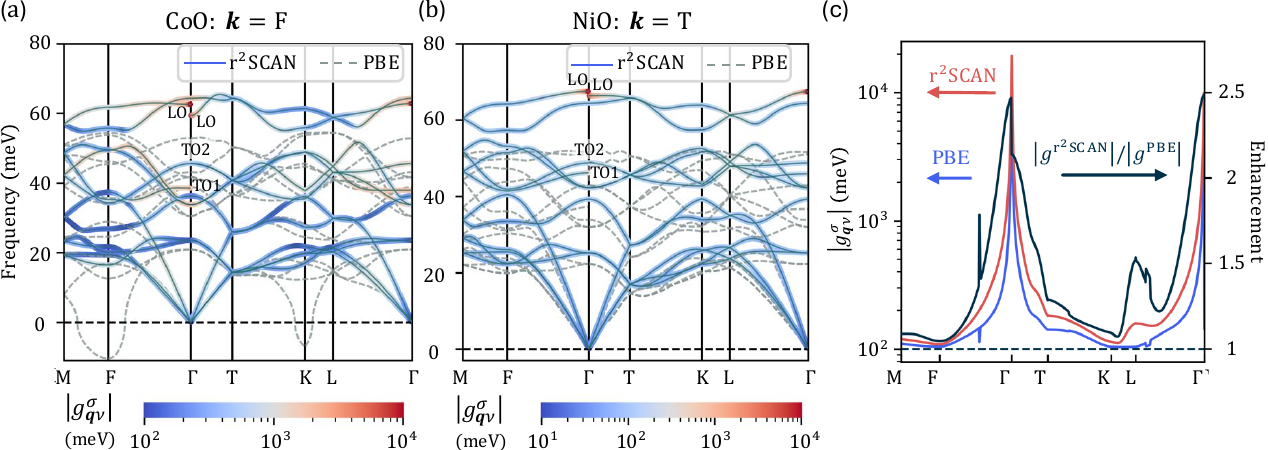}
\caption{\textbf{Calculated phonon dispersions and EPCs in CoO and NiO}. (a) Phonon dispersion in $G$-type AFM CoO along the high-symmetry lines in the BZ, where the LO and TO splittings at the $\Gamma$ point are marked. The color
of markers along the $\mathbf{q}$-path shows the magnitude of EPC matrix elements, calculated by summing over bands ($N_b$) selected to include Co 3$d$ and O 2$p$ orbitals: $|g^\sigma_{\vecq\nu}| \equiv \left( \sum_{mn} |g_{mn,\nu}^\sigma(\veck=K,\vecq)|^2/N_{b} \right)^{1/2}$, where $\sigma$ is chosen to be the spin-up channel, $\vecq$ follows the high-symmetry lines in the bulk BZ, and $\veck$ is fixed at the $K$ point. Note that the PBE EPC matrix elements are not shown here for clarity. 
(b) Same as (a), but for NiO with $\veck=T$. 
(c) The comparison of EPC matrix elements for NiO between PBE and \rrscan. The blue line represents the EPC matrix elements $\vert g_{\vecq\nu}^{\sigma}(\veck=T)\vert$ calculated with the PBE functional, where $\veck$ is fixed at the $T$ point, with $\nu$ denoting the highest optical phonon mode. The red line shows the corresponding data but is calculated using the \rrscan functional. The gray line indicates the ratio, \(\vert g^\text{\rrscan}\vert/\vert g^\text{PBE}\vert\), along the high-symmetry lines in the BZ.}
\label{fig:MO_ph}
\end{figure*}

Here, we apply the \rrscan meta-GGA within the finite-difference approach~\cite{ManuelPRB2020,ManuelPRB2022} to calculate EPCs of transition metal compounds including cobalt oxide (CoO) and nickel oxide (NiO), as well as a main-group conventional superconductor, magnesium diboride (MgB\(_2\)). We show that \rrscan correctly captures strong EPC effects in CoO and NiO, significantly improving over the standard Perdew-Burke-Ernzerhof (PBE) GGA without requiring the introduction of empirical parameters. At the same time, \rrscan achieves high accuracy in modeling phonon-mediated superconducting properties of MgB$_2$.

\SB{Results and Discussion}

\SB{Evaluation of EPCs in CoO and NiO}\textbf{\---}CoO and NiO are two fundamental transition metal oxides that pose challenges for DFT~\cite{Zhang2020b,AlexMOPBEU}. In our simulations, we use the low-temperature antiferromagnetic (AFM) phase of CoO and NiO, characterized by ferromagnetic planes along the [111] direction that alternate in magnetization. Upon relaxation, the cubic rock-salt structures of CoO and NiO exhibit slight distortions due to the AFM order. Experimentally, CoO and NiO are polar semiconductors, which exhibit long-range dipole-dipole interactions arising from the polarization of collective ionic movements. These interactions cause splittings between longitudinal optical (LO) and transverse optical (TO) phonon modes at the $\Gamma$ point—a phenomenon known as LO-TO splitting. To account for this effect, the nonanalytic contributions needed for calculating phonon dispersions and EPCs must be included~\cite{PhononDFPTRMP2001, Ab_phonon_semiconductors}, which depend on the Born effective charges ($Z^{*}$) and the ion-clamped macroscopic dielectric tensor ($\epsilon^\infty$).

Figure~\ref{fig:MO_ph}(a) presents the phonon dispersions of CoO calculated using \rrscan and PBE, along with the EPC matrix elements from \rrscan. PBE predicts unphysical negative phonon frequencies along the $M$-$F$-$\Gamma$ and $T$-$K$-$L$ paths, indicating lattice instability in CoO, as reported in previous studies~\cite{DFPTUCoOCeltech,DFPTUeph}. Consequently, PBE is unsuitable for calculating the EPC in CoO. The PBE+$U$ method is often applied to eliminate these negative frequencies, allowing DFPT+$U$ to provide accurate EPC predictions~\cite{DFPTUCoOCeltech,DFPTUeph,Lattic_dynamics_CoO_DFTU}. Remarkably, \rrscan resolves these lattice instabilities without requiring any $U$ parameters, as shown in Fig.~\ref{fig:MO_ph}(a). Furthermore, a comparison of phonon dispersions calculated by \rrscan and PBE reveals that \rrscan shifts phonon frequencies—particularly for the optical modes—upward, achieving better agreement with experimental results, as shown in Figs.~\ref{fig:MO_ph}(a) and \textcolor{blue}{S2} of the Supplementary Materials (SM).

In Fig.~\ref{fig:MO_ph}(a), our \rrscan calculations also reveal significant LO-TO splitting in CoO. Note that PBE doesn't capture this LO-TO splitting because  
it incorrectly predicts a metallic ground state for CoO [Fig.~S1 (\textcolor{blue}{b})]. 
The predicted values of $\omega_\text{LO}= (59.4,\, 62.6)$ meV and $\omega_\text{TO}= (38.5,\, 45.7)$ meV from \rrscan are close to experimental measurements [see Fig.~\textcolor{blue}{S2}]. Additionally, we observe notable splitting between the TO1 and TO2 modes along the $F-\Gamma-T$ direction [Fig.~\ref{fig:MO_ph}(a)]. This splitting primarily results from the structural distortion induced by the AFM order, which breaks the equivalence along the [111] directions ($\Gamma-T$), as noted in previous studies~\cite{DFPTUCoOCeltech,DFPTUNiOHardy}. A similar, though smaller, LO mode splitting is observed at the $\Gamma$ point. Similar results were obtained using PBE+$U$ calculations~\cite{DFPTUCoOCeltech}, although the top two optical modes along the $F-\Gamma-T$ direction are reversed in our \rrscan results compared to PBE+$U$, likely due to the larger Born effective charge $Z^*$ (2.25) and ion-clamped macroscopic dielectric function $\varepsilon^\infty$ (6.17) values predicted by PBE+$U$~\cite{DFPTUCoOCeltech}.

\begin{table*}[htbp]\setlength{\belowcaptionskip}{0.5cm}
\renewcommand\arraystretch{1.3}
\begin{ruledtabular}
\begin{tabular}{lcccccccccc}
&\multicolumn{5}{c}{CoO} &\multicolumn{5}{c}{NiO} \\
\cline{2-6} \cline{7-11}
&E$_\text{g}$ (eV)   &$Z^{*}$   &\textit{m} ($\mu_B$)   &$\varepsilon^\infty$  &$\varepsilon^\infty_\text{RPA}$
&E$_\text{g}$ (eV)   &$Z^{*}$   &\textit{m} ($\mu_B$)   &$\varepsilon^\infty$  &$\varepsilon^\infty_\text{RPA}$ \\
\colrule
PBE
&0    &--     &2.36   &--      &--
&0.81 &2.38   &1.30   &20.70   &19.67 \\
\rrscan
&0.85 &1.83  &2.58    &5.48   &5.44
&2.97 &2.23  &1.58    &7.52   &7.10 \\
\cline{5-6}\cline{10-11}
Expt.
&2.8~\cite{PR1959_CoO_gap}   &2.06~\cite{PR_CoO_Borncharge}, 1.78~\cite{JAP_borncharge_CoO} &3.8~\cite{CoO_magmoment} &\multicolumn{2}{c}{5.3~\cite{JAP_borncharge_CoO}  }
&4.0~\cite{NiO_expt_gap}, 4.3~\cite{NiO_expt_gap_prl1984}   &2.2~\cite{Zhikun_Liiu_NiO} &1.9~\cite{NiO_expt_magmoment}  &\multicolumn{2}{c}{5.7~\cite{JAP_borncharge_CoO}}
\end{tabular}
\end{ruledtabular}
\caption{The calculated band gap ($E_g$), diagonals of the Born effective charge tensor ($Z^{*}$), local magnetic moments (\textit{m}) and ion-clamped macroscopic dielectric constants in the RPA ($\varepsilon^\infty_\text{RPA}$) and including $f_{xc}$ ($\varepsilon^\infty$) of CoO and NiO for the PBE and \rrscan functionals, as well as measured experimental values. 
}\label{tab:born_charge}
\end{table*}

Figure~\ref{fig:MO_ph}(a) also shows that our \rrscan calculations reveal substantial EPC contributions from both LO and TO modes near the $\Gamma$ point, with maximum EPC matrix elements [\textit{g} defined in Eq.~\eqref{g_element}] exceeding $10^{4}$ meV. This result aligns with previous DFPT+$U$ findings~\cite{DFPTUeph} quantitatively. In transition metal oxides like CoO and NiO, the polar nature of the bonds leads to strong interactions between electrons and LO and TO phonons via the Fröhlich mechanism~\cite{Frolichinteraction}. These interactions become particularly significant as the wave vector $\mathbf{q} \rightarrow 0$, causing a divergence in the EPC matrix elements for the LO and TO modes [Fig.~\ref{fig:MO_ph} (a)]. These interactions are often substantial enough to form large polarons—quasiparticles composed of an electron or hole coupled with lattice distortions—these play a crucial role in charge transport and the dynamic behavior of electrons in polar materials. It is a significant advantage of \rrscan to capture the Fröhlich interaction in polar materials without relying on empirical parameters~\cite{DFPTUeph,STO_ZhouJinJian,GaAs_Zhou}.

The phonon dispersions of NiO calculated using \rrscan and PBE are shown in Fig.~\ref{fig:MO_ph}(b), where a similar phonon softening trend from \rrscan to PBE is observed, as seen in CoO. Additionally, \rrscan accurately captures the LO-TO mode splitting in NiO due to the inclusion of the nonanalytic term. Similar to CoO, \rrscan calculations reveal a clear splitting of the TO1 and TO2 modes at the $\Gamma$ point in NiO, though the LO mode splitting is comparatively smaller, as discussed in Ref.~\cite{DFPTUNiOHardy}. Notably, the TO mode splitting in NiO is smaller than in CoO [Fig.\ref{fig:MO_ph}(b)], which can be attributed to the smaller local magnetic moments of NiO [Table~\ref{tab:born_charge}], as pointed out in the previous study~\cite{DFPTUCoOCeltech}. Importantly, the phonon characteristics computed with \rrscan show strong consistency with those obtained using PBE+$U$, as reported in Ref.~\cite{DFPTUNiOHardy}.

Similar to CoO, strong EPC contributions from the LO and TO modes in NiO are also predicted by \rrscan [Fig.\ref{fig:MO_ph}(b)]. In addition, to illustrate the differences between PBE and \rrscan in EPC predictions, we plot the EPC matrix elements of NiO for the highest optical mode along the high-symmetry $\mathbf{q}$-path for each functional in Fig.\ref{fig:MO_ph}(c). The EPC matrix elements are significantly enhanced by \rrscan compared to PBE, particularly around the $\Gamma$ point, with an approximate increase of 2.5, as shown in Fig.~\ref{fig:MO_ph}(c).

\begin{figure}[htpb]
\centering
\includegraphics[width=0.99\columnwidth]{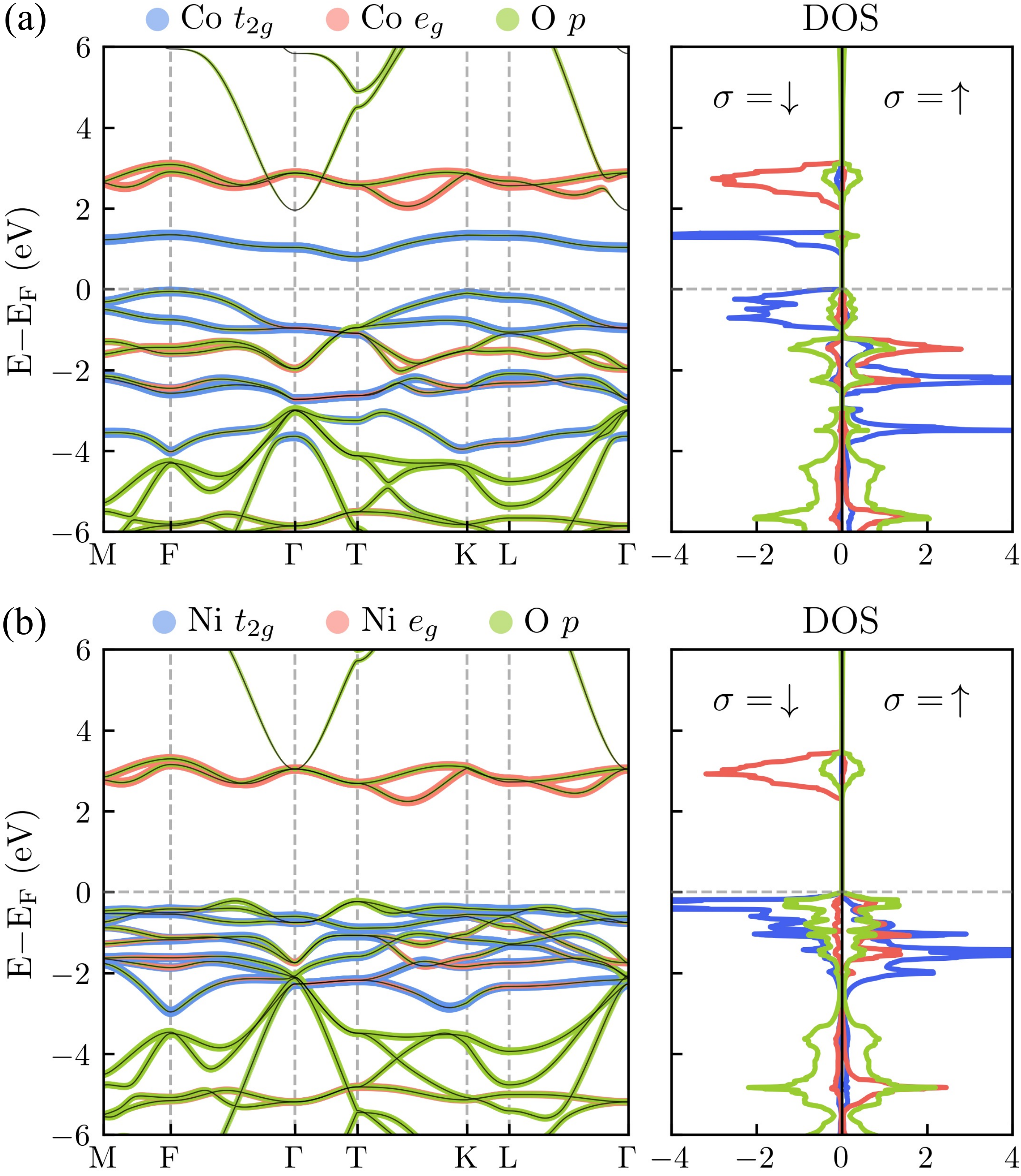}
\caption{\textbf{The calculated electronic structure of $G$-type AFM phase of CoO and NiO using the \rrscan functional}. (a) Orbitally-projected band structure of CoO along high-symmetry lines in the BZ, where the Co $t_{2g}$ and $e_{g}$, and O $p$ orbitals are marked with blue, red, and green colors, respectively. The side panel shows the partial density of states of the respective orbitals. (b) Same as (a), but for NiO.}
\label{fig:MO_el}
\end{figure}

\SB{Why does \rrscan yield superior EPC predictions in CoO and NiO?}\textbf{---} The results above clearly demonstrate that \rrscan significantly improves the accuracy of EPC descriptions for CoO and NiO over PBE. Our analysis indicates that the superior performance of \rrscan over PBE can largely be attributed to the more exact constraints that \rrscan satisfies~\cite{Sun2015,sun2016accurate, R2SCANJPCL}. Additionally, for transition and rare-earth metal compounds, \rrscan exhibits reduced a self-interaction error (SIE) compared to PBE, significantly improving the description of their electronic structures [Fig.~\ref{fig:MO_el} and S1]~\cite{Zhang2020, ZhangR2020}. The significant improvement of \rrscan over PBE in describing the phonon characteristics of NiO, particularly for the optical modes, has also been attributed to the reduction of the SIE~\cite{NingCMPhonon}. SIE arises from the imperfect cancellation of the spurious classical Coulomb self-interaction by the approximate exchange-correlation self-interaction~\cite{perdew1981self}. Since the repulsive self-Coulomb term exceeds the attractive self-exchange-correlation term, the net SIE is generally positive for semi-local functionals, causing orbitals to be underbound (with orbital energies too high) and wave functions to be excessively delocalized. This leads to electron densities that are overly spread out and easily perturbed, ultimately resulting in underestimated magnetic moments and band gaps.
The reduction of the SIE in \rrscan compared to PBE therefore results in less delocalization error, more compact $d$- and $f$-orbitals, and fewer fractional occupations, which stabilize the magnetic moments~\cite{Zhang2020b,ZhangR2020} and increase the band gaps for NiO and CoO as shown in Table~\ref{tab:born_charge}.

Table~\ref{tab:born_charge} also presents the calculated $Z^{*}$ and $\varepsilon^\infty$ values for CoO and NiO from both PBE and \rrscan, compared with experimental data. Both $Z^{*}$ and $\varepsilon^\infty$ are critical quantities used in the nonanalytic contributions needed for calculating phonon dispersions and EPCs. The Born effective charge $Z^{*}$ measures the dynamic charge of an atom in a crystal, capturing how polarization changes in response to the atomic displacement and accounting for the redistribution of electronic charge under an applied electric field \cite{BornCharge_vasp1,BornCharge_vasp2, king1993theory,resta1993macroscopic}. The ion-clamped macroscopic dielectric tensor $\varepsilon^\infty$ describes the long-range dielectric response of a material, considering only the electronic contribution without ionic motion, reflecting how the electronic cloud in a crystal responds to an external electric field \cite{BornCharge_vasp1,BornCharge_vasp2, king1993theory,resta1993macroscopic}. $Z^{*}$ and $\varepsilon^\infty$ are expected to be overestimated due to the SIE that leads to overly delocalized and easily perturbed electron densities. This can be seen by comparing the $Z^*$ and $\varepsilon^\infty$ values calculated from \rrscan and PBE against the experimental data. For CoO, we find that the \rrscan predictions for $Z^*$ and $\varepsilon^\infty$ are in good agreement with experimental results~\cite{PR_CoO_Borncharge,JAP_borncharge_CoO}. However, PBE incorrectly predicts a metallic ground state for CoO as shown in Fig.~{\color{blue}S1}(b), to be discussed later, resulting in the vanishing of the nonanalytic terms. For NiO, PBE significantly overestimates $\varepsilon^\infty$ by a factor of 3.77, whereas \rrscan provides significant improvement, reducing the overestimation to 31.93\%. The calculated $Z^*$ value using \rrscan is 2.23, which is in excellent agreement with the experimental value of 2.2, while PBE overestimates $Z^*$ by 8.18\%. These results corroborate that \rrscan reduces SIE in comparison with PBE for NiO and CoO.

The above analysis strongly indicates that the improvement of \rrscan over PBE on EPC matrix elements originates from the SIE reduction. In the context of the Kohn-Sham (KS) DFT~\cite{FelicianoRMP2017}, the EPC matrix element is defined as:
\begin{equation}\label{g_element}
g_{mn,\nu}({\veck},\vecq) = \langle \psi_{m{\veck+\vecq}} \vert \partial_{\vecq\nu}v^\text{KS} \vert \psi_{n{\veck}}\rangle,
\end{equation}
where $\psi_{n{\veck}}$ is the electronic wavefunction with the momentum $\veck$ in the band $n$, and $\partial_{\vecq\nu}v^\text{KS}$ the derivative of the self-consistent KS potential with respect to a phonon of wavevector $\vecq$, band index $\nu$, and frequency $\omega_{\vecq\nu}$. We note that \(\partial_{\vecq\nu}v^\text{KS}\) is calculated by the finite difference approach here. By definition \(\partial_{\vecq\nu}v^\text{KS}\) is given by
 \begin{equation}
   \partial_{\vecq\nu}v^\text{KS} = \partial_{\vecq\nu}v^\text{en}\ + \partial_{\vecq\nu}v^\text{H}\ +
   \partial_{\vecq\nu}v^\text{xc},\
   \label{eq:ks}
   \end{equation}
where \(v^\text{en}\) is the potential arising from electron-nuclei interactions, \(v^\text{H}\) is the Hartree potential, and \(v^\text{xc}\) is the exchange-correlation  potential. This can be recast into~\cite{Onida_RMP}: 
\begin{equation}\label{vks}
   \partial_{\mathbf{q}\nu} v^\text{KS} =(\varepsilon^\text{Hxc})^{-1} \partial_{\mathbf{q}\nu} v^\text{en}. 
\end{equation}
We have used the definition of the `test electron' dielectric matrix,
\begin{equation}\label{hxc}
  \varepsilon^\text{Hxc} = 1-(v^\text{C}+f^\text{xc})\chi^{0}.
\end{equation}
Here, $v^\text{C}$ is the Coulomb kernel, $f^\text{xc}(\mathbf{r},\mathbf{r}^\prime)\equiv\frac{\delta^2 E_\text{xc}[n]}{\delta n(\mathbf{r}) \delta n(\mathbf{r}^\prime)}$ the exchange-correlation kernel, and  \(\chi^{0}(\mathbf{r},\mathbf{r}^\prime)\equiv\frac{\delta n(\mathbf{r})}{\delta v^\text{KS} (\mathbf{r}^\prime)}\) the KS non-interacting linear response function.

Based on Eqs.~\eqref{vks} and~\eqref{hxc}, one can see that the change in the KS potential due to the motion of the nuclei originates from two contributions: (i) the change in the bare electron-nuclei potential, and (ii) the response of the electronic degrees of freedom to screen these interactions at the DFA level. Thus, improvements of EPC matrix elements within the KS DFT framework should come from two key contributions: $\chi_0$ and $f^\text{xc}$. 

The ion-clamped macroscopic dielectric constants $\varepsilon^\infty$ are quantities related to the head ($\mathbf{q} \rightarrow 0$) of $\varepsilon^\text{Hxc}$~\cite{Linear_optical_paw}. 
For a given density functional, $\varepsilon^\infty$ can be calculated using the Random Phase Approximation (RPA) where $f^\text{xc}$ is ignored, with the value denoted as $\varepsilon^\infty_\text{RPA}$~\cite{Linear_optical_paw,Dielectric_HSE}.  Table~\ref{tab:born_charge} shows that the difference between $\varepsilon^\infty_\text{RPA}$ and $\varepsilon^\infty$ of a density functional is much smaller than the difference between $\varepsilon^\infty$ of different density functionals, suggesting that $f^\text{xc}$ has a less important effect on $\varepsilon^\infty$ than the $\chi_0$ effect of the chosen density functional. We note that \rrscan satisfies more exact constraints than PBE does with more accurate total energies in general, suggesting that \rrscan can have a better $f^\text{xc}$~\cite{Nazarov_exchangekernel}.

The KS linear response function \(\chi_0\) describes how the electron density \(n\) reacts to small changes in the Kohn-Sham potential \(v^\text{KS}\). This linear response function is often overestimated in magnitude by SIE, which leads to an overly delocalized and easily perturbable electron density, as well as an underestimated band gap that is inversely proportional to the linear response function.

For CoO, PBE fails to open a band gap, as illustrated in Fig.~\textcolor{blue}{S1}(b). This leads to too large $\chi_0$, resulting in dynamical instabilities with the appearance of imaginary frequencies in the phonon calculations as shown in Figure~\ref{fig:MO_ph}(a). In contrast, due to the SIE reduction, \rrscan predicts an indirect band gap of 0.85 eV, attributed to the splitting between the occupied $t_{2g}$ orbitals and the unoccupied $t_{2g}$ orbitals in the spin-down channel, as shown in Fig.~\ref{fig:MO_el}(a). This undoubtedly improves $\chi_0$ and thus provides an improved phonon dispersion as well as strong EPC effects at the LO and TO phonon modes around the $\Gamma$ point as shown in Figure~\ref{fig:MO_ph}(a). Since a smaller gap generally correlates with a larger $\varepsilon^\infty$ \cite{penn1962wave,onishi2024universal}, the improved gaps from \rrscan explain our values of $\varepsilon^\infty$ being closer to experimental results as shown in Table.~\ref{tab:born_charge}.

For NiO, \rrscan predicts an indirect band gap of 2.97 eV, with the valence band maximum (VBM) at the $T$ point and the conduction band minimum (CBM) along the $T$-$K$ direction, significantly larger than the 0.8 eV band gap predicted by PBE [Fig.~\textcolor{blue}{S1}(d)] but still noticeably smaller than the 4.0$\sim$4.3 eV of experimental results [see Table~\ref{tab:born_charge}]. PBE significantly underestimates the band gap, yielding a larger $\chi_0$, which, in turn, produces smaller EPC matrix elements compared to \rrscan, as shown in Fig.~\ref{fig:MO_ph}(c).

From the above analysis, accurate EPC predictions can be expected within the KS-DFT framework if the SIE in the exchange-correlation approximation is minimized. This can be achieved by advancing up Jacob's ladder of DFT~\cite{Jacob_ladder_DFT}, moving from LDA to GGA, to meta-GGA, and then to hybrid density functionals, PZ-SIC \cite{perdew1981self}, etc., or by applying empirical corrections to the chosen functional, such as DFT+$U$~\cite{LDAU_first,LDAU_Sawatzky}. The HSE hybrid density functional~\cite{heyd2003hybrid} has become a standard for calculating band gaps in solids due to its reduced SIE; however, its computational cost—one or two orders of magnitude higher than LDA, GGAs, and metaGGAs—limits its practicality for EPC calculations. DFT+$U$ offers an alternative to metaGGA functionals like \rrscan, but the empirical nature of the $U$ parameters reduces its predictive power~\cite{LDAU_first,LDAU_Sawatzky}.

\SB{Performance of \rrscan in conventional superconductor MgB$_2$}\textbf{\---}Having examined the improvements that \rrscan offers for predicting EPCs in complex compounds like CoO and NiO with $d$-electrons, we now demonstrate that \rrscan also provides reliable predictions of the EPC strength $\lambda$ [defined in Eq.~\eqref{eq:lam}] in main group compounds. Here, we revisit the prototypical conventional superconductor MgB$_2$, a well-known phonon-mediated multiband superconductor with a critical temperature ($T_\text{c}$) of approximately 39 K at ambient pressure~\cite{MgB2_Nagamatsu2001,Twogap_MgB2,MgB2_MGPRL}. MgB$_2$ has been extensively studied, both experimentally~\cite{MgB2_Nagamatsu2001,Twogap_MgB2,MgB2_MGPRL,Kogn_anomaly_MgB2,WeakanharmonicMgB2} and theoretically~\cite{LouieMgB2nature,Anisotropic_ME,Kortus_MgB2,Giant_Anharmon_MgB2,Choi_PRB_Louie}.

\begin{figure*}[ht]\centering
\includegraphics[width=\textwidth]{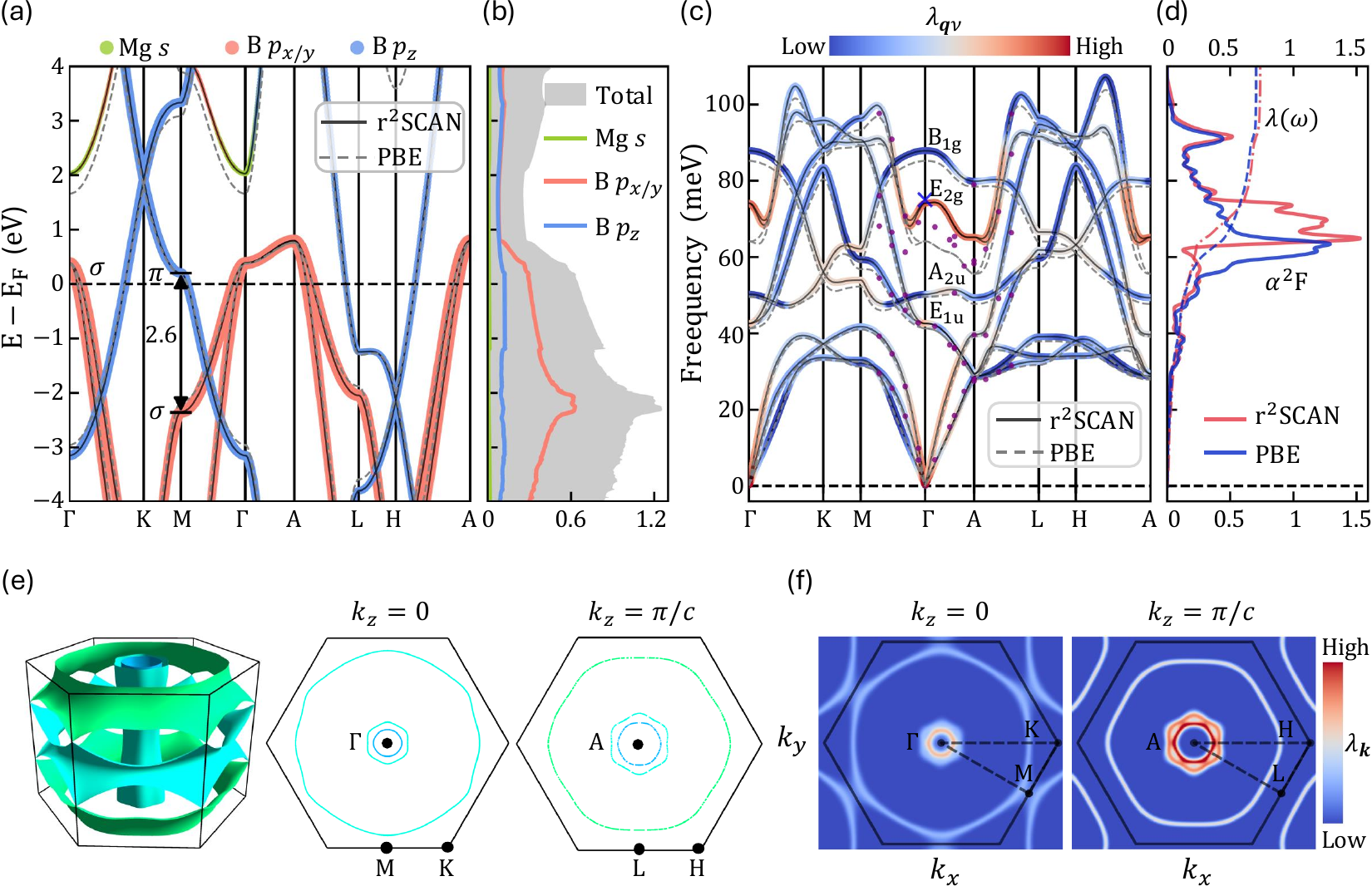}  % Adjust the file name as needed
\caption{\textbf{Calculated electronic structure, phonon dispersion and EPC of MgB$_2$.}
(a) The calculated band structure of MgB$_2$ along high-symmetry lines in the BZ from \rrscan (solid black lines) and PBE (dash gray lines). Orbital characters from \rrscan are also shown, where Mg $s$, B $p_{x/y}$, and B $p_{z}$ orbitals are indicated in green, red, and blue, respectively. The splitting of $\pi$ and $\sigma$ bands from \rrscan at the $M$ point is highlighted. Orbital projections from PBE are omitted for clarity.
(b) The total and partial density of states for each orbital using \rrscan.
(c) Phonon dispersion along high-symmetry lines in the BZ, comparing \rrscan (solid black lines) and PBE (dashed gray lines). Marker color intensity corresponds to the magnitude of the EPC strength, $\lambda_{\textbf{q}\nu}$ from \rrscan calculations. EPC strength $\lambda_{\textbf{q}\nu}$ values from PBE calculations are not shown. Experimental phonon data from inelastic X-ray scattering (magenta solid points), extracted from Ref.~\cite{WeakanharmonicMgB2}, are also included for comparison. The measured frequency of the $E_{2g}$ mode at the $\Gamma$ point from Raman spectroscopy is 75 meV~\cite{WeakanharmonicMgB2,Raman_mgB2_prb,Raman_mgB2_pressure}, represented by a blue cross.
(d) Calculated Eliashberg spectral functions $\alpha^2F(\omega)$ and cumulative EPC strength $\lambda$ for MgB$_2$, using \rrscan (red lines) and PBE (green lines).  A \textit{k}-mesh of $100\times100\times100$ and a \textit{q}-mesh of $50\times50\times50$ were used for the BZ sampling.
(e) Left panel: The calculated three-dimensional Fermi surface of MgB$_2$ obtained using the Wannier model based on \rrscan calculations with a $200\times200\times200$ \textit{k}-mesh. Middle and right panels: The Fermi surface projections on the $k_z = 0$ and $k_z = \pi/c$ planes, respectively.
(f) Distribution of $k$-resolved EPC strength $\lambda_{\veck} = \sum_{mn,\nu\vecq}|g_{mn,\nu}({\veck},\vecq)|^{2}\delta(\varepsilon_{m\textbf{k}+\textbf{q}}-\varepsilon_\text{F} )\delta(\varepsilon_{n\textbf{k}}-\varepsilon_\text{F} )\times(2/\omega_{\textbf{q}\nu})$ from \rrscan across the $k_x - k_y$ plane at $k_z = 0$ (left panel) and $k_z = \pi/c$ (right panel).}
\label{fig:mgb2}
\end{figure*}

MgB$_2$ crystallizes in the tetragonal Bravais lattice with space group $P6/mmm$ (\#191, 6/mmm), where boron atoms form honeycomb lattices separated by hexagonal layers of magnesium atoms [Fig. {\color{blue}S3}]. The optimized lattice parameters of MgB$_2$ using \rrscan are $a = 3.063$ \AA{} and $c = 3.521$ \AA{}, which are in excellent agreement with the corresponding experimental values $a= 3.086$ \AA{} and $c = 3.524$ \AA{}~\cite{MgB2_Nagamatsu2001}. 
The electronic states at the Fermi level calculated from \rrscan are primarily derived from boron $p$-states. Specifically, two $\sigma$-bonding bands, contributed by B-$p_{x/y}$ orbitals, cross the Fermi level [Fig.~\ref{fig:mgb2}(a)], forming two small cylindrical hole Fermi surfaces along the $\Gamma-A$ direction [Fig.~\ref{fig:mgb2}(e)]. These $\sigma$ bands are confined to the boron planes, exhibiting a two-dimensional character. In contrast, the dispersive $\pi$-bonding bands, derived from the B-$p_z$ orbitals, exhibit a three-dimensional character [Fig.~\ref{fig:mgb2}(e)], resulting in one hole and one electron Fermi surfaces near the $k_{z} = 0$ and the $k_{z} = \pm \pi/c$ planes, respectively. Our findings are generally consistent with previous calculations using standard DFT~\cite{Kortus_MgB2,Giant_Anharmon_MgB2}. However, subtle differences between our \rrscan results and those obtained using PBE are noticeable, as illustrated in Fig.~\ref{fig:mgb2} (a). For instance, the bandwidth predicted by \rrscan is generally larger than that predicted by PBE. At the $M$ point, the band splitting between the upper $\pi$ band and the lower $\sigma$ band is approximately 2.6 eV with \rrscan, compared to 2.4 eV with PBE. Notably, the \rrscan prediction matches the experimental value of 2.6 eV~\cite{Anisotropic_optical_MgB2}, which also agrees with high-level calculations such as modified HSE of 2.6 eV~\cite{ZhipingPRX2013} and scQPGW of 2.6 eV~\cite{ZhipingPRX2013}.

Figure~\ref{fig:mgb2}(c) presents the calculated phonon dispersion of MgB$_2$ using the \rrscan and PBE functionals. Consistent with prior studies~\cite{WeakanharmonicMgB2,Giant_Anharmon_MgB2}, \rrscan identifies four optical phonon modes at the $\Gamma$ point:  $E_{1u}$ ,  $A_{2u}$,  $E_{2g}$, and  $B_{1g}$, with frequencies of 42.6, 50.2, 74.1, and 87.9 meV, respectively, in close agreement with experimental values~\cite{WeakanharmonicMgB2}. Notably, the $E_{2g}$ mode, involving in-plane stretching of B atoms, is critical for EPCs in MgB$_2$~\cite{Choi_PRB_Louie,Anisotropic_ME,LouieMgB2nature}, as we can see in Figs.~\ref{fig:mgb2} (c) and (f). This comparison between \rrscan and PBE results [Fig.~\ref{fig:mgb2}(c)] shows excellent agreement for acoustic modes, while \rrscan provides a higher and more accurate prediction for the $E_{2g}$ mode frequency of 74.1 meV, matching the experimental value of 75 meV obtained from Raman measurements~\cite{WeakanharmonicMgB2,Raman_mgB2_prb,Raman_mgB2_pressure}. PBE predicts a lower value of 63 meV for the $E_{2g}$ mode. This discrepancy with the experimental value is often attributed to the overestimated bond lengths by PBE, resulting in softer optical phonon modes. While previous studies have linked the difference between PBE predictions and experimental values to anharmonic effects~\cite{Giant_Anharmon_MgB2,Choi_PRB_Louie}, recent findings suggest that MgB$_2$ exhibits only weak anharmonicity~\cite{MgB2_phexpt}.  This suggests that the improved results by \rrscan arise from a better description of exchange-correlation effects.

Figure~\ref{fig:mgb2}(d) shows the calculated isotropic Eliashberg function $\alpha^{2}F (\omega)$ of MgB$_2$, based on Eq.~\eqref{eq:a2F}, using the \rrscan (red lines) and PBE (blue lines) functionals. Our PBE results reproduce the dominant peak around 60 meV, primarily contributed by the E$_{2g}$ mode, consistent with previous findings~\cite{Anisotropic_ME}. A second peak appears around 91 meV, slightly higher than previous results obtained with the LDA functional~\cite{Anisotropic_ME}. The calculated $\alpha^2F(\omega)$ using the PBE functional yields an EPC strength $\lambda$  of approximately 0.70, close to the earlier value of 0.74 reported for the LDA functional~\cite{Anisotropic_ME}. The \rrscan results, shown as a solid red line, exhibit similarities to the PBE results in the range of 0–50 meV, since the predicted phonon dispersions from both functionals are nearly identical for low-frequency phonons [Fig.~\ref{fig:mgb2} (c)]. However, the dominant peaks from the \rrscan calculations appear in the range of 63–75 meV, which are much broader than those from the PBE results, reflecting the larger phonon bandwidth of the $E_{2g}$ mode predicted by \rrscan. The second peak at 91 meV is also captured in our \rrscan results.
These results also reflect the softening of the optical phonon modes in PBE. The broader distribution of the dominant peak in $\alpha^{2}F(\omega)$ observed with \rrscan results in an EPC strength $\lambda$ of 0.73, slightly larger than the PBE value of 0.70. According to Eqs.~\eqref{eq:tc} and \eqref{wlog}, the estimated $\omega_{log}$ is $\sim$58.1 meV, and the predicted $T_\text{c}$ ranges from 22.4$\sim$30.3 K, with the adjustable parameter $\mu^{*}$ set between 0.12 and 0.08, consistent with previous calculations~\cite{Anisotropic_ME}.

Based on the above analysis, our results demonstrate that the \rrscan functional provides a reliable description of electronic structure, phonon dispersion, and EPC in MgB$_2$ compared to PBE. While previous studies have shown that MgB$_2$ is a multi-gap superconductor~\cite{MgB2_MGPRL,Twogap_MgB2,LouieMgB2nature,Choi_PRB_Louie,EPW2007, Anisotropic_ME,FlorisSCDFT}, solving the superconducting gap equation is beyond the scope of the present work.

\SB{Conclusion}\textbf{\---}{Calculating EPCs for complex materials with $d$- and $f$-electrons has long been a challenge for first-principles methods due to the need for both accuracy and efficiency. In this study, we demonstrate that the efficient \rrscan density functional significantly improves EPC predictions compared to the standard PBE density functional for transition-metal compounds such as CoO and NiO, alongside better descriptions of their electronic structures and phonon dispersions. Due to SIE, PBE incorrectly predicts CoO as metallic with imaginary phonon frequencies, rendering it unsuitable for EPC calculations. In contrast, \rrscan reduces SIE, opens the band gap, stabilizes the phonon dispersion, and predicts strong EPCs for optical modes. Analysis of the dielectric screening tensors for CoO and NiO also indicates that these improvements stem from the reduction of SIE in \rrscan compared to PBE.

Additionally, for the main-group compound MgB$_2$, \rrscan provides improved predictions of the band structure and phonon dispersion compared to PBE. Notably, \rrscan predicts a phonon frequency for the critical E$_{2g}$ mode that closely matches Raman measurements, whereas PBE significantly underestimates this value. The resulting EPC strength $\lambda$ and isotropic Eliashberg spectral function $\alpha^2F(\omega)$ calculated using \rrscan show good agreement with prior studies.

Our findings highlight the general efficacy of the \rrscan functional in accurately predicting EPCs across a wide range of materials. The study underscores the potential of advanced functionals like \rrscan to enhance the predictive power of first-principles EPC calculations, enabling deeper and more accurate insights into the physical properties of materials with strong electron-electron and electron-phonon interactions.
}

\SB{Methods}\textbf{\---}First-principles calculations were performed using the projector-augmented-wave (PAW) method~\cite{Kresse1999}, as implemented in the Vienna ab initio simulation package (VASP)~\cite{Kresse1993,Kresse1996}. Both the \rrscan and PBE density functionals were employed in our calculations. For calculating the EPC, we follow the Wannier-interpolation procedure outlined in Ref.~\cite{ManuelPRB2020} while using the all-electron EPC matrix elements defined in Ref.~\cite{ManuelPRB2022,FDEPCPAW}. The Wannier representation of the electronic band structure was obtained using Wannier90~\cite{MOSTOFI20142309} with the help of the VASP2WANNIER90 interface. We calculate the EPC strength $\lambda$ and the isotropic Eliashberg spectral function $\alpha^{2}F$ based on the following equations.

Within the Migdal approximation~\cite{Migdal1958}, the imaginary part of the phonon self-energy is expressed as
\begin{equation}\label{eq:phse}
\Pi''_{\vecq\nu} = \text{Im} \sum_{m,n,\veck} w_{\veck}  \vert g_{mn,\nu}( {\veck,\vecq} ) \vert^2 
\frac{ f_{n\veck} - f_{m{\veck+\vecq}} }{
\epsilon_{m{\veck+\vecq}} - \epsilon_{n\veck} - \omega_{\vecq\nu} - i\eta},
\end{equation}
where $f_{n{\veck}}$ is the Fermi occupation of the single-particle state $\vert\psi_{n\veck}\rangle$ associated with the eigenvalue $\epsilon_{n\veck}$, $w_{\veck}$ are the weights of the dense ${\veck}$-points, and $\eta$ indicates a real positive infinitesimal. We note here that the spin index is omitted for simplicity. The EPC strength $\lambda$ is the full BZ average of the mode-resolved $\lambda_{\vecq\nu}$
\begin{equation}\label{eq:lam}
\lambda = \sum_{\vecq\nu} w_{\vecq} \lambda_{\vecq\nu},
\end{equation}
with $w_{\vecq}$ being the weights of the dense $\vecq$-points. In our implementations, the double delta approximation~\cite{epiq_impl2024} is adopted to calculate the imaginary part of the phonon self-energy, under which the Eq.~\eqref{eq:phse} can be rewritten as follows:

\begin{equation}\label{eq:phse_doubledelta}
\Pi''_{\vecq\nu} = \text{Im} \sum_{m,n,\veck} w_{\veck}  \vert g_{mn,\nu}( {\veck,\vecq} ) \vert^2 \delta(\epsilon_{n{\bf k}})\delta(\epsilon_{m{\bf k}+{\bf q}})
\end{equation}
Under the double delta approximation, the $\lambda_{\vecq\nu}$ can be expressed as:
\begin{equation}
\lambda_{\vecq\nu} = \frac{1}{\pi N_\text{F}} \frac{ \Pi''_{\vecq\nu}}{ \omega^2_{\vecq\nu}},
\end{equation}
where $N_\text{F}$ is the density of states at the Fermi level. Using this mode-resolved $\lambda_{\vecq\nu}$ and the phonon frequencies $\omega_{\vecq\nu}$, the isotropic Eliashberg spectral function, $\alpha^2 F$, can be calculated by
\begin{equation}\label{eq:a2F}
\alpha^2F(\omega) = \frac{1}{2}\sum_{\vecq\nu}  
w_{\vecq} \omega_{\vecq\nu} \lambda_{\vecq\nu} \, \delta( \omega - \omega_{\vecq\nu}).
\end{equation}

The phonon-mediated superconducting transition temperature, $T_\text{c}$, is evaluated with the McMillian-Allen-Dynes formula~\cite{McMillan1968},
\begin{equation}
T_\text{c} = \frac{\omega_\text{log}}{1.2}\exp\left[{\frac{-1.04(1+\lambda)}{\lambda(1 - 0.62\mu^*) - \mu^*}}\right],
\label{eq:tc}
\end{equation}
where the logarithmic average frequency $\omega_{\text{log}}$ can be calculated from the following equation~\cite{Allen1975,grimvall1981electron}
\begin{equation}\label{wlog}
\omega_\text{log} = \exp\left[\frac{2}{\lambda}\int^\infty_0\text{d}\omega \frac{ \log\,\omega}{\omega}\alpha^2F(\omega)\right].
\end{equation}

\textit{CoO and NiO}\textbf{\---} All the calculations were performed within the density functional theory framework with projector-augmented wave potentials~\cite{Kresse1999} using the VASP~\cite{Kresse1993,Kresse1996}. An initial $G$-type antiferromagnetic (AFM) structure with 4 atoms was used for structural relaxation. A high-energy cutoff of 600 eV was applied to truncate the plane-wave basis set. Crystal structures and ionic positions were fully optimized with a force convergence criterion of 0.001 eV/\AA{} for each atom and a total energy tolerance of $10^{-8}$ eV. A $\Gamma$-centered k-point mesh of $15 \times 15 \times 15$ was employed for the structural relaxation. For phonon calculations, a $4 \times 4 \times 4$ supercell with a $\Gamma$-centered $k$-point mesh of $3 \times 3 \times 3$ was used, employing the finite displacement method. A Gaussian smearing with a small $\sigma$ of 0.01 was applied in our phonon calculations. Transition metal (Ni and Co) $d$ states and oxygen $p$ states were included in constructing the Wannier functions.

\textit{MgB$_2$}\textbf{\---}The calculations were performed within the density functional theory framework with projector-augmented wave potentials~\cite{Kresse1999} using the VASP~\cite{Kresse1993,Kresse1996}. A high-energy cutoff of 520 eV was applied to truncate the plane-wave basis set. Crystal structures and ionic positions were fully optimized using a force convergence criterion of 0.001 eV/\AA{} for each atom, along with a total energy tolerance of $10^{-8}$ eV. To relax the structures, we adopted a $\Gamma$-centered k-point mesh of $13 \times 13 \times 12$ to sample the primitive bulk BZ. For phonon calculations, a $6 \times 6 \times 6$ supercell with a $\Gamma$-centered k-point mesh of $7 \times 7 \times 9$ was employed using the finite displacement method. A Gaussian smearing with a small $\sigma$ of 0.01 was applied in our phonon calculations. Mg $p$ and $s$ states, along with B $p$ and $s$ states, were included in constructing the Wannier functions.

\bibliography{Ref}

\begin{thebibliography}{86}%
\makeatletter
\providecommand \@ifxundefined [1]{%
 \@ifx{#1\undefined}
}%
\providecommand \@ifnum [1]{%
 \ifnum #1\expandafter \@firstoftwo
 \else \expandafter \@secondoftwo
 \fi
}%
\providecommand \@ifx [1]{%
 \ifx #1\expandafter \@firstoftwo
 \else \expandafter \@secondoftwo
 \fi
}%
\providecommand \natexlab [1]{#1}%
\providecommand \enquote  [1]{``#1''}%
\providecommand \bibnamefont  [1]{#1}%
\providecommand \bibfnamefont [1]{#1}%
\providecommand \citenamefont [1]{#1}%
\providecommand \href@noop [0]{\@secondoftwo}%
\providecommand \href [0]{\begingroup \@sanitize@url \@href}%
\providecommand \@href[1]{\@@startlink{#1}\@@href}%
\providecommand \@@href[1]{\endgroup#1\@@endlink}%
\providecommand \@sanitize@url [0]{\catcode `\\12\catcode `\$12\catcode
  `\&12\catcode `\#12\catcode `\^12\catcode `\_12\catcode `\%12\relax}%
\providecommand \@@startlink[1]{}%
\providecommand \@@endlink[0]{}%
\providecommand \url  [0]{\begingroup\@sanitize@url \@url }%
\providecommand \@url [1]{\endgroup\@href {#1}{\urlprefix }}%
\providecommand \urlprefix  [0]{URL }%
\providecommand \Eprint [0]{\href }%
\providecommand \doibase [0]{http://dx.doi.org/}%
\providecommand \selectlanguage [0]{\@gobble}%
\providecommand \bibinfo  [0]{\@secondoftwo}%
\providecommand \bibfield  [0]{\@secondoftwo}%
\providecommand \translation [1]{[#1]}%
\providecommand \BibitemOpen [0]{}%
\providecommand \bibitemStop [0]{}%
\providecommand \bibitemNoStop [0]{.\EOS\space}%
\providecommand \EOS [0]{\spacefactor3000\relax}%
\providecommand \BibitemShut  [1]{\csname bibitem#1\endcsname}%
\let\auto@bib@innerbib\@empty
%</preamble>
\bibitem [{\citenamefont {Giustino}(2017)}]{FelicianoRMP2017}%
  \BibitemOpen
  \bibfield  {author} {\bibinfo {author} {\bibfnamefont {F.}~\bibnamefont
  {Giustino}},\ }\href {\doibase 10.1103/RevModPhys.89.015003} {\bibfield
  {journal} {\bibinfo  {journal} {Rev. Mod. Phys.}\ }\textbf {\bibinfo {volume}
  {89}},\ \bibinfo {pages} {015003} (\bibinfo {year} {2017})}\BibitemShut
  {NoStop}%
\bibitem [{\citenamefont {Grimvall}(1981)}]{grimvall1981electron}%
  \BibitemOpen
  \bibfield  {author} {\bibinfo {author} {\bibfnamefont {G.}~\bibnamefont
  {Grimvall}},\ }\href@noop {} {\emph {\bibinfo {title} {The Electron-Phonon
  Interaction in Metals}}},\ \bibinfo {edition} {1st}\ ed.\ (\bibinfo
  {publisher} {North-Holland Publishing Co. Amsterdam, New York, Oxford},\
  \bibinfo {address} {Amsterdam},\ \bibinfo {year} {1981})\ p.\ \bibinfo
  {pages} {304}\BibitemShut {NoStop}%
\bibitem [{\citenamefont {Pellegrini}\ and\ \citenamefont
  {Sanna}(2024)}]{Pellegrini2024NRP}%
  \BibitemOpen
  \bibfield  {author} {\bibinfo {author} {\bibfnamefont {C.}~\bibnamefont
  {Pellegrini}}\ and\ \bibinfo {author} {\bibfnamefont {A.}~\bibnamefont
  {Sanna}},\ }\href {\doibase 10.1038/s42254-024-00738-9} {\bibfield  {journal}
  {\bibinfo  {journal} {Nature Reviews Physics}\ }\textbf {\bibinfo {volume}
  {6}},\ \bibinfo {pages} {509} (\bibinfo {year} {2024})}\BibitemShut {NoStop}%
\bibitem [{\citenamefont {Engel}\ \emph {et~al.}(2020)\citenamefont {Engel},
  \citenamefont {Marsman}, \citenamefont {Franchini},\ and\ \citenamefont
  {Kresse}}]{ManuelPRB2020}%
  \BibitemOpen
  \bibfield  {author} {\bibinfo {author} {\bibfnamefont {M.}~\bibnamefont
  {Engel}}, \bibinfo {author} {\bibfnamefont {M.}~\bibnamefont {Marsman}},
  \bibinfo {author} {\bibfnamefont {C.}~\bibnamefont {Franchini}}, \ and\
  \bibinfo {author} {\bibfnamefont {G.}~\bibnamefont {Kresse}},\ }\href
  {\doibase 10.1103/PhysRevB.101.184302} {\bibfield  {journal} {\bibinfo
  {journal} {Phys. Rev. B}\ }\textbf {\bibinfo {volume} {101}},\ \bibinfo
  {pages} {184302} (\bibinfo {year} {2020})}\BibitemShut {NoStop}%
\bibitem [{\citenamefont {Engel}\ \emph {et~al.}(2022)\citenamefont {Engel},
  \citenamefont {Miranda}, \citenamefont {Chaput}, \citenamefont {Togo},
  \citenamefont {Verdi}, \citenamefont {Marsman},\ and\ \citenamefont
  {Kresse}}]{ManuelPRB2022}%
  \BibitemOpen
  \bibfield  {author} {\bibinfo {author} {\bibfnamefont {M.}~\bibnamefont
  {Engel}}, \bibinfo {author} {\bibfnamefont {H.}~\bibnamefont {Miranda}},
  \bibinfo {author} {\bibfnamefont {L.}~\bibnamefont {Chaput}}, \bibinfo
  {author} {\bibfnamefont {A.}~\bibnamefont {Togo}}, \bibinfo {author}
  {\bibfnamefont {C.}~\bibnamefont {Verdi}}, \bibinfo {author} {\bibfnamefont
  {M.}~\bibnamefont {Marsman}}, \ and\ \bibinfo {author} {\bibfnamefont
  {G.}~\bibnamefont {Kresse}},\ }\href {\doibase 10.1103/PhysRevB.106.094316}
  {\bibfield  {journal} {\bibinfo  {journal} {Phys. Rev. B}\ }\textbf {\bibinfo
  {volume} {106}},\ \bibinfo {pages} {094316} (\bibinfo {year}
  {2022})}\BibitemShut {NoStop}%
\bibitem [{\citenamefont {Baroni}\ \emph {et~al.}(2001)\citenamefont {Baroni},
  \citenamefont {de~Gironcoli}, \citenamefont {Dal~Corso},\ and\ \citenamefont
  {Giannozzi}}]{PhononDFPTRMP2001}%
  \BibitemOpen
  \bibfield  {author} {\bibinfo {author} {\bibfnamefont {S.}~\bibnamefont
  {Baroni}}, \bibinfo {author} {\bibfnamefont {S.}~\bibnamefont
  {de~Gironcoli}}, \bibinfo {author} {\bibfnamefont {A.}~\bibnamefont
  {Dal~Corso}}, \ and\ \bibinfo {author} {\bibfnamefont {P.}~\bibnamefont
  {Giannozzi}},\ }\href {\doibase 10.1103/RevModPhys.73.515} {\bibfield
  {journal} {\bibinfo  {journal} {Rev. Mod. Phys.}\ }\textbf {\bibinfo {volume}
  {73}},\ \bibinfo {pages} {515} (\bibinfo {year} {2001})}\BibitemShut
  {NoStop}%
\bibitem [{\citenamefont {Shang}\ and\ \citenamefont
  {Yang}(2023)}]{ShanghJCPeph}%
  \BibitemOpen
  \bibfield  {author} {\bibinfo {author} {\bibfnamefont {H.}~\bibnamefont
  {Shang}}\ and\ \bibinfo {author} {\bibfnamefont {J.}~\bibnamefont {Yang}},\
  }\href {\doibase 10.1063/5.0140724} {\bibfield  {journal} {\bibinfo
  {journal} {The Journal of Chemical Physics}\ }\textbf {\bibinfo {volume}
  {158}},\ \bibinfo {pages} {130901} (\bibinfo {year} {2023})}\BibitemShut
  {NoStop}%
\bibitem [{\citenamefont {Monserrat}(2018)}]{Monserrat2018FD}%
  \BibitemOpen
  \bibfield  {author} {\bibinfo {author} {\bibfnamefont {B.}~\bibnamefont
  {Monserrat}},\ }\href {\doibase 10.1088/1361-648X/aaa737} {\bibfield
  {journal} {\bibinfo  {journal} {Journal of Physics: Condensed Matter}\
  }\textbf {\bibinfo {volume} {30}},\ \bibinfo {pages} {083001} (\bibinfo
  {year} {2018})}\BibitemShut {NoStop}%
\bibitem [{\citenamefont {Giustino}\ \emph {et~al.}(2007)\citenamefont
  {Giustino}, \citenamefont {Cohen},\ and\ \citenamefont {Louie}}]{EPW2007}%
  \BibitemOpen
  \bibfield  {author} {\bibinfo {author} {\bibfnamefont {F.}~\bibnamefont
  {Giustino}}, \bibinfo {author} {\bibfnamefont {M.~L.}\ \bibnamefont {Cohen}},
  \ and\ \bibinfo {author} {\bibfnamefont {S.~G.}\ \bibnamefont {Louie}},\
  }\href {\doibase 10.1103/PhysRevB.76.165108} {\bibfield  {journal} {\bibinfo
  {journal} {Phys. Rev. B}\ }\textbf {\bibinfo {volume} {76}},\ \bibinfo
  {pages} {165108} (\bibinfo {year} {2007})}\BibitemShut {NoStop}%
\bibitem [{\citenamefont {Zhou}\ \emph
  {et~al.}(2021{\natexlab{a}})\citenamefont {Zhou}, \citenamefont {Park},
  \citenamefont {Lu}, \citenamefont {Maliyov}, \citenamefont {Tong},\ and\
  \citenamefont {Bernardi}}]{Perturbo}%
  \BibitemOpen
  \bibfield  {author} {\bibinfo {author} {\bibfnamefont {J.-J.}\ \bibnamefont
  {Zhou}}, \bibinfo {author} {\bibfnamefont {J.}~\bibnamefont {Park}}, \bibinfo
  {author} {\bibfnamefont {I.-T.}\ \bibnamefont {Lu}}, \bibinfo {author}
  {\bibfnamefont {I.}~\bibnamefont {Maliyov}}, \bibinfo {author} {\bibfnamefont
  {X.}~\bibnamefont {Tong}}, \ and\ \bibinfo {author} {\bibfnamefont
  {M.}~\bibnamefont {Bernardi}},\ }\href {\doibase
  https://doi.org/10.1016/j.cpc.2021.107970} {\bibfield  {journal} {\bibinfo
  {journal} {Computer Physics Communications}\ }\textbf {\bibinfo {volume}
  {264}},\ \bibinfo {pages} {107970} (\bibinfo {year}
  {2021}{\natexlab{a}})}\BibitemShut {NoStop}%
\bibitem [{\citenamefont {Antonius}\ \emph {et~al.}(2014)\citenamefont
  {Antonius}, \citenamefont {Ponc\'e}, \citenamefont {Boulanger}, \citenamefont
  {C\^ot\'e},\ and\ \citenamefont {Gonze}}]{FDGWdiamond}%
  \BibitemOpen
  \bibfield  {author} {\bibinfo {author} {\bibfnamefont {G.}~\bibnamefont
  {Antonius}}, \bibinfo {author} {\bibfnamefont {S.}~\bibnamefont {Ponc\'e}},
  \bibinfo {author} {\bibfnamefont {P.}~\bibnamefont {Boulanger}}, \bibinfo
  {author} {\bibfnamefont {M.}~\bibnamefont {C\^ot\'e}}, \ and\ \bibinfo
  {author} {\bibfnamefont {X.}~\bibnamefont {Gonze}},\ }\href {\doibase
  10.1103/PhysRevLett.112.215501} {\bibfield  {journal} {\bibinfo  {journal}
  {Phys. Rev. Lett.}\ }\textbf {\bibinfo {volume} {112}},\ \bibinfo {pages}
  {215501} (\bibinfo {year} {2014})}\BibitemShut {NoStop}%
\bibitem [{\citenamefont {Yin}\ \emph {et~al.}(2013)\citenamefont {Yin},
  \citenamefont {Kutepov},\ and\ \citenamefont {Kotliar}}]{ZhipingPRX2013}%
  \BibitemOpen
  \bibfield  {author} {\bibinfo {author} {\bibfnamefont {Z.~P.}\ \bibnamefont
  {Yin}}, \bibinfo {author} {\bibfnamefont {A.}~\bibnamefont {Kutepov}}, \ and\
  \bibinfo {author} {\bibfnamefont {G.}~\bibnamefont {Kotliar}},\ }\href
  {\doibase 10.1103/PhysRevX.3.021011} {\bibfield  {journal} {\bibinfo
  {journal} {Phys. Rev. X}\ }\textbf {\bibinfo {volume} {3}},\ \bibinfo {pages}
  {021011} (\bibinfo {year} {2013})}\BibitemShut {NoStop}%
\bibitem [{\citenamefont {Furness}\ \emph {et~al.}(2018)\citenamefont
  {Furness}, \citenamefont {Zhang}, \citenamefont {Lane}, \citenamefont {Buda},
  \citenamefont {Barbiellini}, \citenamefont {Markiewicz}, \citenamefont
  {Bansil},\ and\ \citenamefont {Sun}}]{Furness2018}%
  \BibitemOpen
  \bibfield  {author} {\bibinfo {author} {\bibfnamefont {J.~W.}\ \bibnamefont
  {Furness}}, \bibinfo {author} {\bibfnamefont {Y.}~\bibnamefont {Zhang}},
  \bibinfo {author} {\bibfnamefont {C.}~\bibnamefont {Lane}}, \bibinfo {author}
  {\bibfnamefont {I.~G.}\ \bibnamefont {Buda}}, \bibinfo {author}
  {\bibfnamefont {B.}~\bibnamefont {Barbiellini}}, \bibinfo {author}
  {\bibfnamefont {R.~S.}\ \bibnamefont {Markiewicz}}, \bibinfo {author}
  {\bibfnamefont {A.}~\bibnamefont {Bansil}}, \ and\ \bibinfo {author}
  {\bibfnamefont {J.}~\bibnamefont {Sun}},\ }\href {\doibase
  10.1038/s42005-018-0009-4} {\bibfield  {journal} {\bibinfo  {journal}
  {Communications Physics}\ }\textbf {\bibinfo {volume} {1}},\ \bibinfo {pages}
  {11} (\bibinfo {year} {2018})}\BibitemShut {NoStop}%
\bibitem [{\citenamefont {Lane}\ \emph {et~al.}(2018)\citenamefont {Lane},
  \citenamefont {Furness}, \citenamefont {Buda}, \citenamefont {Zhang},
  \citenamefont {Markiewicz}, \citenamefont {Barbiellini}, \citenamefont
  {Sun},\ and\ \citenamefont {Bansil}}]{Lane2018}%
  \BibitemOpen
  \bibfield  {author} {\bibinfo {author} {\bibfnamefont {C.}~\bibnamefont
  {Lane}}, \bibinfo {author} {\bibfnamefont {J.~W.}\ \bibnamefont {Furness}},
  \bibinfo {author} {\bibfnamefont {I.~G.}\ \bibnamefont {Buda}}, \bibinfo
  {author} {\bibfnamefont {Y.}~\bibnamefont {Zhang}}, \bibinfo {author}
  {\bibfnamefont {R.~S.}\ \bibnamefont {Markiewicz}}, \bibinfo {author}
  {\bibfnamefont {B.}~\bibnamefont {Barbiellini}}, \bibinfo {author}
  {\bibfnamefont {J.}~\bibnamefont {Sun}}, \ and\ \bibinfo {author}
  {\bibfnamefont {A.}~\bibnamefont {Bansil}},\ }\href {\doibase
  10.1103/PhysRevB.98.125140} {\bibfield  {journal} {\bibinfo  {journal}
  {Physical Review B}\ }\textbf {\bibinfo {volume} {98}},\ \bibinfo {pages}
  {125140} (\bibinfo {year} {2018})}\BibitemShut {NoStop}%
\bibitem [{\citenamefont {Zhang}\ \emph
  {et~al.}(2020{\natexlab{a}})\citenamefont {Zhang}, \citenamefont {Furness},
  \citenamefont {Zhang}, \citenamefont {Wang}, \citenamefont {Zunger},\ and\
  \citenamefont {Sun}}]{Zhang2020b}%
  \BibitemOpen
  \bibfield  {author} {\bibinfo {author} {\bibfnamefont {Y.}~\bibnamefont
  {Zhang}}, \bibinfo {author} {\bibfnamefont {J.}~\bibnamefont {Furness}},
  \bibinfo {author} {\bibfnamefont {R.}~\bibnamefont {Zhang}}, \bibinfo
  {author} {\bibfnamefont {Z.}~\bibnamefont {Wang}}, \bibinfo {author}
  {\bibfnamefont {A.}~\bibnamefont {Zunger}}, \ and\ \bibinfo {author}
  {\bibfnamefont {J.}~\bibnamefont {Sun}},\ }\href {\doibase
  10.1103/PhysRevB.102.045112} {\bibfield  {journal} {\bibinfo  {journal}
  {Phys. Rev. B}\ }\textbf {\bibinfo {volume} {102}},\ \bibinfo {pages}
  {045112} (\bibinfo {year} {2020}{\natexlab{a}})}\BibitemShut {NoStop}%
\bibitem [{\citenamefont {Zhang}\ \emph {et~al.}(2024)\citenamefont {Zhang},
  \citenamefont {Lane}, \citenamefont {Nokelainen}, \citenamefont {Singh},
  \citenamefont {Barbiellini}, \citenamefont {Markiewicz}, \citenamefont
  {Bansil},\ and\ \citenamefont {Sun}}]{ZhangPRLLaNiO2}%
  \BibitemOpen
  \bibfield  {author} {\bibinfo {author} {\bibfnamefont {R.}~\bibnamefont
  {Zhang}}, \bibinfo {author} {\bibfnamefont {C.}~\bibnamefont {Lane}},
  \bibinfo {author} {\bibfnamefont {J.}~\bibnamefont {Nokelainen}}, \bibinfo
  {author} {\bibfnamefont {B.}~\bibnamefont {Singh}}, \bibinfo {author}
  {\bibfnamefont {B.}~\bibnamefont {Barbiellini}}, \bibinfo {author}
  {\bibfnamefont {R.~S.}\ \bibnamefont {Markiewicz}}, \bibinfo {author}
  {\bibfnamefont {A.}~\bibnamefont {Bansil}}, \ and\ \bibinfo {author}
  {\bibfnamefont {J.}~\bibnamefont {Sun}},\ }\href {\doibase
  10.1103/PhysRevLett.133.066401} {\bibfield  {journal} {\bibinfo  {journal}
  {Phys. Rev. Lett.}\ }\textbf {\bibinfo {volume} {133}},\ \bibinfo {pages}
  {066401} (\bibinfo {year} {2024})}\BibitemShut {NoStop}%
\bibitem [{\citenamefont {Ning}\ \emph {et~al.}(2024)\citenamefont {Ning},
  \citenamefont {Lane}, \citenamefont {Barbiellini}, \citenamefont
  {Markiewicz}, \citenamefont {Bansil}, \citenamefont {Ruzsinszky},
  \citenamefont {Perdew},\ and\ \citenamefont {Sun}}]{NingJCPDFAYBCO6}%
  \BibitemOpen
  \bibfield  {author} {\bibinfo {author} {\bibfnamefont {J.}~\bibnamefont
  {Ning}}, \bibinfo {author} {\bibfnamefont {C.}~\bibnamefont {Lane}}, \bibinfo
  {author} {\bibfnamefont {B.}~\bibnamefont {Barbiellini}}, \bibinfo {author}
  {\bibfnamefont {R.~S.}\ \bibnamefont {Markiewicz}}, \bibinfo {author}
  {\bibfnamefont {A.}~\bibnamefont {Bansil}}, \bibinfo {author} {\bibfnamefont
  {A.}~\bibnamefont {Ruzsinszky}}, \bibinfo {author} {\bibfnamefont {J.~P.}\
  \bibnamefont {Perdew}}, \ and\ \bibinfo {author} {\bibfnamefont
  {J.}~\bibnamefont {Sun}},\ }\href {\doibase 10.1063/5.0181349} {\bibfield
  {journal} {\bibinfo  {journal} {The Journal of Chemical Physics}\ }\textbf
  {\bibinfo {volume} {160}},\ \bibinfo {pages} {064106} (\bibinfo {year}
  {2024})}\BibitemShut {NoStop}%
\bibitem [{\citenamefont {Ning}\ \emph {et~al.}(2022)\citenamefont {Ning},
  \citenamefont {Furness},\ and\ \citenamefont {Sun}}]{NingCMPhonon}%
  \BibitemOpen
  \bibfield  {author} {\bibinfo {author} {\bibfnamefont {J.}~\bibnamefont
  {Ning}}, \bibinfo {author} {\bibfnamefont {J.~W.}\ \bibnamefont {Furness}}, \
  and\ \bibinfo {author} {\bibfnamefont {J.}~\bibnamefont {Sun}},\ }\href
  {\doibase 10.1021/acs.chemmater.1c03222} {\bibfield  {journal} {\bibinfo
  {journal} {Chemistry of Materials}\ }\textbf {\bibinfo {volume} {34}},\
  \bibinfo {pages} {2562} (\bibinfo {year} {2022})}\BibitemShut {NoStop}%
\bibitem [{\citenamefont {Ning}\ \emph {et~al.}(2023)\citenamefont {Ning},
  \citenamefont {Lane}, \citenamefont {Zhang}, \citenamefont {Matzelle},
  \citenamefont {Singh}, \citenamefont {Barbiellini}, \citenamefont
  {Markiewicz}, \citenamefont {Bansil},\ and\ \citenamefont
  {Sun}}]{NingPRBYBCO6}%
  \BibitemOpen
  \bibfield  {author} {\bibinfo {author} {\bibfnamefont {J.}~\bibnamefont
  {Ning}}, \bibinfo {author} {\bibfnamefont {C.}~\bibnamefont {Lane}}, \bibinfo
  {author} {\bibfnamefont {Y.}~\bibnamefont {Zhang}}, \bibinfo {author}
  {\bibfnamefont {M.}~\bibnamefont {Matzelle}}, \bibinfo {author}
  {\bibfnamefont {B.}~\bibnamefont {Singh}}, \bibinfo {author} {\bibfnamefont
  {B.}~\bibnamefont {Barbiellini}}, \bibinfo {author} {\bibfnamefont {R.~S.}\
  \bibnamefont {Markiewicz}}, \bibinfo {author} {\bibfnamefont
  {A.}~\bibnamefont {Bansil}}, \ and\ \bibinfo {author} {\bibfnamefont
  {J.}~\bibnamefont {Sun}},\ }\href {\doibase 10.1103/PhysRevB.107.045126}
  {\bibfield  {journal} {\bibinfo  {journal} {Phys. Rev. B}\ }\textbf {\bibinfo
  {volume} {107}},\ \bibinfo {pages} {045126} (\bibinfo {year}
  {2023})}\BibitemShut {NoStop}%
\bibitem [{\citenamefont {Zhou}\ \emph
  {et~al.}(2021{\natexlab{b}})\citenamefont {Zhou}, \citenamefont {Park},
  \citenamefont {Timrov}, \citenamefont {Floris}, \citenamefont {Cococcioni},
  \citenamefont {Marzari},\ and\ \citenamefont {Bernardi}}]{DFPTUeph}%
  \BibitemOpen
  \bibfield  {author} {\bibinfo {author} {\bibfnamefont {J.-J.}\ \bibnamefont
  {Zhou}}, \bibinfo {author} {\bibfnamefont {J.}~\bibnamefont {Park}}, \bibinfo
  {author} {\bibfnamefont {I.}~\bibnamefont {Timrov}}, \bibinfo {author}
  {\bibfnamefont {A.}~\bibnamefont {Floris}}, \bibinfo {author} {\bibfnamefont
  {M.}~\bibnamefont {Cococcioni}}, \bibinfo {author} {\bibfnamefont
  {N.}~\bibnamefont {Marzari}}, \ and\ \bibinfo {author} {\bibfnamefont
  {M.}~\bibnamefont {Bernardi}},\ }\href {\doibase
  10.1103/PhysRevLett.127.126404} {\bibfield  {journal} {\bibinfo  {journal}
  {Phys. Rev. Lett.}\ }\textbf {\bibinfo {volume} {127}},\ \bibinfo {pages}
  {126404} (\bibinfo {year} {2021}{\natexlab{b}})}\BibitemShut {NoStop}%
\bibitem [{\citenamefont {Floris}\ \emph {et~al.}(2020)\citenamefont {Floris},
  \citenamefont {Timrov}, \citenamefont {Himmetoglu}, \citenamefont {Marzari},
  \citenamefont {de~Gironcoli},\ and\ \citenamefont
  {Cococcioni}}]{DFPTUCoOCeltech}%
  \BibitemOpen
  \bibfield  {author} {\bibinfo {author} {\bibfnamefont {A.}~\bibnamefont
  {Floris}}, \bibinfo {author} {\bibfnamefont {I.}~\bibnamefont {Timrov}},
  \bibinfo {author} {\bibfnamefont {B.}~\bibnamefont {Himmetoglu}}, \bibinfo
  {author} {\bibfnamefont {N.}~\bibnamefont {Marzari}}, \bibinfo {author}
  {\bibfnamefont {S.}~\bibnamefont {de~Gironcoli}}, \ and\ \bibinfo {author}
  {\bibfnamefont {M.}~\bibnamefont {Cococcioni}},\ }\href {\doibase
  10.1103/PhysRevB.101.064305} {\bibfield  {journal} {\bibinfo  {journal}
  {Phys. Rev. B}\ }\textbf {\bibinfo {volume} {101}},\ \bibinfo {pages}
  {064305} (\bibinfo {year} {2020})}\BibitemShut {NoStop}%
\bibitem [{\citenamefont {Li}\ \emph {et~al.}(2019)\citenamefont {Li},
  \citenamefont {Antonius}, \citenamefont {Wu}, \citenamefont {da~Jornada},\
  and\ \citenamefont {Louie}}]{ZhengLuGWPTBaTiO3}%
  \BibitemOpen
  \bibfield  {author} {\bibinfo {author} {\bibfnamefont {Z.}~\bibnamefont
  {Li}}, \bibinfo {author} {\bibfnamefont {G.}~\bibnamefont {Antonius}},
  \bibinfo {author} {\bibfnamefont {M.}~\bibnamefont {Wu}}, \bibinfo {author}
  {\bibfnamefont {F.~H.}\ \bibnamefont {da~Jornada}}, \ and\ \bibinfo {author}
  {\bibfnamefont {S.~G.}\ \bibnamefont {Louie}},\ }\href {\doibase
  10.1103/PhysRevLett.122.186402} {\bibfield  {journal} {\bibinfo  {journal}
  {Phys. Rev. Lett.}\ }\textbf {\bibinfo {volume} {122}},\ \bibinfo {pages}
  {186402} (\bibinfo {year} {2019})}\BibitemShut {NoStop}%
\bibitem [{\citenamefont {Li}\ \emph {et~al.}(2021)\citenamefont {Li},
  \citenamefont {Wu}, \citenamefont {Chan},\ and\ \citenamefont
  {Louie}}]{ZhengluGWPTCuprate}%
  \BibitemOpen
  \bibfield  {author} {\bibinfo {author} {\bibfnamefont {Z.}~\bibnamefont
  {Li}}, \bibinfo {author} {\bibfnamefont {M.}~\bibnamefont {Wu}}, \bibinfo
  {author} {\bibfnamefont {Y.-H.}\ \bibnamefont {Chan}}, \ and\ \bibinfo
  {author} {\bibfnamefont {S.~G.}\ \bibnamefont {Louie}},\ }\href {\doibase
  10.1103/PhysRevLett.126.146401} {\bibfield  {journal} {\bibinfo  {journal}
  {Phys. Rev. Lett.}\ }\textbf {\bibinfo {volume} {126}},\ \bibinfo {pages}
  {146401} (\bibinfo {year} {2021})}\BibitemShut {NoStop}%
\bibitem [{\citenamefont {Sun}\ \emph {et~al.}(2015)\citenamefont {Sun},
  \citenamefont {Ruzsinszky},\ and\ \citenamefont {Perdew}}]{Sun2015}%
  \BibitemOpen
  \bibfield  {author} {\bibinfo {author} {\bibfnamefont {J.}~\bibnamefont
  {Sun}}, \bibinfo {author} {\bibfnamefont {A.}~\bibnamefont {Ruzsinszky}}, \
  and\ \bibinfo {author} {\bibfnamefont {J.}~\bibnamefont {Perdew}},\ }\href
  {\doibase 10.1103/PhysRevLett.115.036402} {\bibfield  {journal} {\bibinfo
  {journal} {Physical Review Letters}\ }\textbf {\bibinfo {volume} {115}},\
  \bibinfo {pages} {036402} (\bibinfo {year} {2015})}\BibitemShut {NoStop}%
\bibitem [{\citenamefont {Furness}\ \emph {et~al.}(2020)\citenamefont
  {Furness}, \citenamefont {Kaplan}, \citenamefont {Ning}, \citenamefont
  {Perdew},\ and\ \citenamefont {Sun}}]{R2SCANJPCL}%
  \BibitemOpen
  \bibfield  {author} {\bibinfo {author} {\bibfnamefont {J.~W.}\ \bibnamefont
  {Furness}}, \bibinfo {author} {\bibfnamefont {A.~D.}\ \bibnamefont {Kaplan}},
  \bibinfo {author} {\bibfnamefont {J.}~\bibnamefont {Ning}}, \bibinfo {author}
  {\bibfnamefont {J.~P.}\ \bibnamefont {Perdew}}, \ and\ \bibinfo {author}
  {\bibfnamefont {J.}~\bibnamefont {Sun}},\ }\href {\doibase
  10.1021/acs.jpclett.0c02405} {\bibfield  {journal} {\bibinfo  {journal} {The
  Journal of Physical Chemistry Letters}\ }\textbf {\bibinfo {volume} {11}},\
  \bibinfo {pages} {8208} (\bibinfo {year} {2020})}\BibitemShut {NoStop}%
\bibitem [{\citenamefont {Sun}\ \emph {et~al.}(2016)\citenamefont {Sun},
  \citenamefont {Remsing}, \citenamefont {Zhang}, \citenamefont {Sun},
  \citenamefont {Ruzsinszky}, \citenamefont {Peng}, \citenamefont {Yang},
  \citenamefont {Paul}, \citenamefont {Waghmare}, \citenamefont {Wu} \emph
  {et~al.}}]{sun2016accurate}%
  \BibitemOpen
  \bibfield  {author} {\bibinfo {author} {\bibfnamefont {J.}~\bibnamefont
  {Sun}}, \bibinfo {author} {\bibfnamefont {R.~C.}\ \bibnamefont {Remsing}},
  \bibinfo {author} {\bibfnamefont {Y.}~\bibnamefont {Zhang}}, \bibinfo
  {author} {\bibfnamefont {Z.}~\bibnamefont {Sun}}, \bibinfo {author}
  {\bibfnamefont {A.}~\bibnamefont {Ruzsinszky}}, \bibinfo {author}
  {\bibfnamefont {H.}~\bibnamefont {Peng}}, \bibinfo {author} {\bibfnamefont
  {Z.}~\bibnamefont {Yang}}, \bibinfo {author} {\bibfnamefont {A.}~\bibnamefont
  {Paul}}, \bibinfo {author} {\bibfnamefont {U.}~\bibnamefont {Waghmare}},
  \bibinfo {author} {\bibfnamefont {X.}~\bibnamefont {Wu}},  \emph {et~al.},\
  }\href@noop {} {\bibfield  {journal} {\bibinfo  {journal} {Nature chemistry}\
  }\textbf {\bibinfo {volume} {8}},\ \bibinfo {pages} {831} (\bibinfo {year}
  {2016})}\BibitemShut {NoStop}%
\bibitem [{\citenamefont {Chen}\ \emph {et~al.}(2017)\citenamefont {Chen},
  \citenamefont {Ko}, \citenamefont {Remsing}, \citenamefont {Andrade},
  \citenamefont {Santra}, \citenamefont {Sun}, \citenamefont {Selloni},
  \citenamefont {Car}, \citenamefont {Klein}, \citenamefont {Perdew} \emph
  {et~al.}}]{chen2017ab}%
  \BibitemOpen
  \bibfield  {author} {\bibinfo {author} {\bibfnamefont {M.}~\bibnamefont
  {Chen}}, \bibinfo {author} {\bibfnamefont {H.-Y.}\ \bibnamefont {Ko}},
  \bibinfo {author} {\bibfnamefont {R.~C.}\ \bibnamefont {Remsing}}, \bibinfo
  {author} {\bibfnamefont {M.~F.~C.}\ \bibnamefont {Andrade}}, \bibinfo
  {author} {\bibfnamefont {B.}~\bibnamefont {Santra}}, \bibinfo {author}
  {\bibfnamefont {Z.}~\bibnamefont {Sun}}, \bibinfo {author} {\bibfnamefont
  {A.}~\bibnamefont {Selloni}}, \bibinfo {author} {\bibfnamefont
  {R.}~\bibnamefont {Car}}, \bibinfo {author} {\bibfnamefont {M.~L.}\
  \bibnamefont {Klein}}, \bibinfo {author} {\bibfnamefont {J.~P.}\ \bibnamefont
  {Perdew}},  \emph {et~al.},\ }\href@noop {} {\bibfield  {journal} {\bibinfo
  {journal} {Proceedings of the National Academy of Sciences}\ }\textbf
  {\bibinfo {volume} {114}},\ \bibinfo {pages} {10846} (\bibinfo {year}
  {2017})}\BibitemShut {NoStop}%
\bibitem [{\citenamefont {Gautam}\ and\ \citenamefont
  {Carter}(2018)}]{gautam2018evaluating}%
  \BibitemOpen
  \bibfield  {author} {\bibinfo {author} {\bibfnamefont {G.~S.}\ \bibnamefont
  {Gautam}}\ and\ \bibinfo {author} {\bibfnamefont {E.~A.}\ \bibnamefont
  {Carter}},\ }\href@noop {} {\bibfield  {journal} {\bibinfo  {journal}
  {Physical Review Materials}\ }\textbf {\bibinfo {volume} {2}},\ \bibinfo
  {pages} {095401} (\bibinfo {year} {2018})}\BibitemShut {NoStop}%
\bibitem [{\citenamefont {Zhang}\ \emph {et~al.}(2022)\citenamefont {Zhang},
  \citenamefont {Singh}, \citenamefont {Lane}, \citenamefont {Kidd},
  \citenamefont {Zhang}, \citenamefont {Barbiellini}, \citenamefont
  {Markiewicz}, \citenamefont {Bansil},\ and\ \citenamefont
  {Sun}}]{ZhangR2020}%
  \BibitemOpen
  \bibfield  {author} {\bibinfo {author} {\bibfnamefont {R.}~\bibnamefont
  {Zhang}}, \bibinfo {author} {\bibfnamefont {B.}~\bibnamefont {Singh}},
  \bibinfo {author} {\bibfnamefont {C.}~\bibnamefont {Lane}}, \bibinfo {author}
  {\bibfnamefont {J.}~\bibnamefont {Kidd}}, \bibinfo {author} {\bibfnamefont
  {Y.}~\bibnamefont {Zhang}}, \bibinfo {author} {\bibfnamefont
  {B.}~\bibnamefont {Barbiellini}}, \bibinfo {author} {\bibfnamefont {R.~S.}\
  \bibnamefont {Markiewicz}}, \bibinfo {author} {\bibfnamefont
  {A.}~\bibnamefont {Bansil}}, \ and\ \bibinfo {author} {\bibfnamefont
  {J.}~\bibnamefont {Sun}},\ }\href {\doibase 10.1103/PhysRevB.105.195134}
  {\bibfield  {journal} {\bibinfo  {journal} {Phys. Rev. B}\ }\textbf {\bibinfo
  {volume} {105}},\ \bibinfo {pages} {195134} (\bibinfo {year}
  {2022})}\BibitemShut {NoStop}%
\bibitem [{\citenamefont {Zhang}\ \emph
  {et~al.}(2020{\natexlab{b}})\citenamefont {Zhang}, \citenamefont {Lane},
  \citenamefont {Furness}, \citenamefont {Barbiellini}, \citenamefont {Perdew},
  \citenamefont {Markiewicz}, \citenamefont {Bansil},\ and\ \citenamefont
  {Sun}}]{Zhang2020}%
  \BibitemOpen
  \bibfield  {author} {\bibinfo {author} {\bibfnamefont {Y.}~\bibnamefont
  {Zhang}}, \bibinfo {author} {\bibfnamefont {C.}~\bibnamefont {Lane}},
  \bibinfo {author} {\bibfnamefont {J.~W.}\ \bibnamefont {Furness}}, \bibinfo
  {author} {\bibfnamefont {B.}~\bibnamefont {Barbiellini}}, \bibinfo {author}
  {\bibfnamefont {J.~P.}\ \bibnamefont {Perdew}}, \bibinfo {author}
  {\bibfnamefont {R.~S.}\ \bibnamefont {Markiewicz}}, \bibinfo {author}
  {\bibfnamefont {A.}~\bibnamefont {Bansil}}, \ and\ \bibinfo {author}
  {\bibfnamefont {J.}~\bibnamefont {Sun}},\ }\href {\doibase
  10.1073/pnas.1910411116} {\bibfield  {journal} {\bibinfo  {journal}
  {Proceedings of the National Academy of Sciences}\ }\textbf {\bibinfo
  {volume} {117}},\ \bibinfo {pages} {68} (\bibinfo {year}
  {2020}{\natexlab{b}})}\BibitemShut {NoStop}%
\bibitem [{\citenamefont {Zhang}\ \emph {et~al.}(2021)\citenamefont {Zhang},
  \citenamefont {Lane}, \citenamefont {Singh}, \citenamefont {Nokelainen},
  \citenamefont {Barbiellini}, \citenamefont {Markiewicz}, \citenamefont
  {Bansil},\ and\ \citenamefont {Sun}}]{Zhang2021}%
  \BibitemOpen
  \bibfield  {author} {\bibinfo {author} {\bibfnamefont {R.}~\bibnamefont
  {Zhang}}, \bibinfo {author} {\bibfnamefont {C.}~\bibnamefont {Lane}},
  \bibinfo {author} {\bibfnamefont {B.}~\bibnamefont {Singh}}, \bibinfo
  {author} {\bibfnamefont {J.}~\bibnamefont {Nokelainen}}, \bibinfo {author}
  {\bibfnamefont {B.}~\bibnamefont {Barbiellini}}, \bibinfo {author}
  {\bibfnamefont {R.~S.}\ \bibnamefont {Markiewicz}}, \bibinfo {author}
  {\bibfnamefont {A.}~\bibnamefont {Bansil}}, \ and\ \bibinfo {author}
  {\bibfnamefont {J.}~\bibnamefont {Sun}},\ }\href {\doibase
  10.1038/s42005-021-00621-4} {\bibfield  {journal} {\bibinfo  {journal}
  {Communications Physics}\ }\textbf {\bibinfo {volume} {4}},\ \bibinfo {pages}
  {118} (\bibinfo {year} {2021})}\BibitemShut {NoStop}%
\bibitem [{\citenamefont {Lane}\ \emph {et~al.}(2023)\citenamefont {Lane},
  \citenamefont {Zhang}, \citenamefont {Barbiellini}, \citenamefont
  {Markiewicz}, \citenamefont {Bansil}, \citenamefont {Sun},\ and\
  \citenamefont {Zhu}}]{Lane2023}%
  \BibitemOpen
  \bibfield  {author} {\bibinfo {author} {\bibfnamefont {C.}~\bibnamefont
  {Lane}}, \bibinfo {author} {\bibfnamefont {R.}~\bibnamefont {Zhang}},
  \bibinfo {author} {\bibfnamefont {B.}~\bibnamefont {Barbiellini}}, \bibinfo
  {author} {\bibfnamefont {R.~S.}\ \bibnamefont {Markiewicz}}, \bibinfo
  {author} {\bibfnamefont {A.}~\bibnamefont {Bansil}}, \bibinfo {author}
  {\bibfnamefont {J.}~\bibnamefont {Sun}}, \ and\ \bibinfo {author}
  {\bibfnamefont {J.-X.}\ \bibnamefont {Zhu}},\ }\href {\doibase
  10.1038/s42005-023-01213-0} {\bibfield  {journal} {\bibinfo  {journal}
  {Communications Physics}\ }\textbf {\bibinfo {volume} {6}},\ \bibinfo {pages}
  {90} (\bibinfo {year} {2023})}\BibitemShut {NoStop}%
\bibitem [{\citenamefont {Trimarchi}\ \emph {et~al.}(2018)\citenamefont
  {Trimarchi}, \citenamefont {Wang},\ and\ \citenamefont
  {Zunger}}]{AlexMOPBEU}%
  \BibitemOpen
  \bibfield  {author} {\bibinfo {author} {\bibfnamefont {G.}~\bibnamefont
  {Trimarchi}}, \bibinfo {author} {\bibfnamefont {Z.}~\bibnamefont {Wang}}, \
  and\ \bibinfo {author} {\bibfnamefont {A.}~\bibnamefont {Zunger}},\ }\href
  {\doibase 10.1103/PhysRevB.97.035107} {\bibfield  {journal} {\bibinfo
  {journal} {Phys. Rev. B}\ }\textbf {\bibinfo {volume} {97}},\ \bibinfo
  {pages} {035107} (\bibinfo {year} {2018})}\BibitemShut {NoStop}%
\bibitem [{\citenamefont {Giannozzi}\ \emph {et~al.}(1991)\citenamefont
  {Giannozzi}, \citenamefont {de~Gironcoli}, \citenamefont {Pavone},\ and\
  \citenamefont {Baroni}}]{Ab_phonon_semiconductors}%
  \BibitemOpen
  \bibfield  {author} {\bibinfo {author} {\bibfnamefont {P.}~\bibnamefont
  {Giannozzi}}, \bibinfo {author} {\bibfnamefont {S.}~\bibnamefont
  {de~Gironcoli}}, \bibinfo {author} {\bibfnamefont {P.}~\bibnamefont
  {Pavone}}, \ and\ \bibinfo {author} {\bibfnamefont {S.}~\bibnamefont
  {Baroni}},\ }\href {\doibase 10.1103/PhysRevB.43.7231} {\bibfield  {journal}
  {\bibinfo  {journal} {Phys. Rev. B}\ }\textbf {\bibinfo {volume} {43}},\
  \bibinfo {pages} {7231} (\bibinfo {year} {1991})}\BibitemShut {NoStop}%
\bibitem [{\citenamefont {Wdowik}\ and\ \citenamefont
  {Parlinski}(2007)}]{Lattic_dynamics_CoO_DFTU}%
  \BibitemOpen
  \bibfield  {author} {\bibinfo {author} {\bibfnamefont {U.~D.}\ \bibnamefont
  {Wdowik}}\ and\ \bibinfo {author} {\bibfnamefont {K.}~\bibnamefont
  {Parlinski}},\ }\href {\doibase 10.1103/PhysRevB.75.104306} {\bibfield
  {journal} {\bibinfo  {journal} {Phys. Rev. B}\ }\textbf {\bibinfo {volume}
  {75}},\ \bibinfo {pages} {104306} (\bibinfo {year} {2007})}\BibitemShut
  {NoStop}%
\bibitem [{\citenamefont {Floris}\ \emph {et~al.}(2011)\citenamefont {Floris},
  \citenamefont {de~Gironcoli}, \citenamefont {Gross},\ and\ \citenamefont
  {Cococcioni}}]{DFPTUNiOHardy}%
  \BibitemOpen
  \bibfield  {author} {\bibinfo {author} {\bibfnamefont {A.}~\bibnamefont
  {Floris}}, \bibinfo {author} {\bibfnamefont {S.}~\bibnamefont
  {de~Gironcoli}}, \bibinfo {author} {\bibfnamefont {E.~K.~U.}\ \bibnamefont
  {Gross}}, \ and\ \bibinfo {author} {\bibfnamefont {M.}~\bibnamefont
  {Cococcioni}},\ }\href {\doibase 10.1103/PhysRevB.84.161102} {\bibfield
  {journal} {\bibinfo  {journal} {Phys. Rev. B}\ }\textbf {\bibinfo {volume}
  {84}},\ \bibinfo {pages} {161102} (\bibinfo {year} {2011})}\BibitemShut
  {NoStop}%
\bibitem [{\citenamefont {Pratt}\ and\ \citenamefont
  {Coelho}(1959)}]{PR1959_CoO_gap}%
  \BibitemOpen
  \bibfield  {author} {\bibinfo {author} {\bibfnamefont {G.~W.}\ \bibnamefont
  {Pratt}}\ and\ \bibinfo {author} {\bibfnamefont {R.}~\bibnamefont {Coelho}},\
  }\href {\doibase 10.1103/PhysRev.116.281} {\bibfield  {journal} {\bibinfo
  {journal} {Phys. Rev.}\ }\textbf {\bibinfo {volume} {116}},\ \bibinfo {pages}
  {281} (\bibinfo {year} {1959})}\BibitemShut {NoStop}%
\bibitem [{\citenamefont {Sakurai}\ \emph {et~al.}(1968)\citenamefont
  {Sakurai}, \citenamefont {Buyers}, \citenamefont {Cowley},\ and\
  \citenamefont {Dolling}}]{PR_CoO_Borncharge}%
  \BibitemOpen
  \bibfield  {author} {\bibinfo {author} {\bibfnamefont {J.}~\bibnamefont
  {Sakurai}}, \bibinfo {author} {\bibfnamefont {W.~J.~L.}\ \bibnamefont
  {Buyers}}, \bibinfo {author} {\bibfnamefont {R.~A.}\ \bibnamefont {Cowley}},
  \ and\ \bibinfo {author} {\bibfnamefont {G.}~\bibnamefont {Dolling}},\ }\href
  {\doibase 10.1103/PhysRev.167.510} {\bibfield  {journal} {\bibinfo  {journal}
  {Phys. Rev.}\ }\textbf {\bibinfo {volume} {167}},\ \bibinfo {pages} {510}
  (\bibinfo {year} {1968})}\BibitemShut {NoStop}%
\bibitem [{\citenamefont {Gielisse}\ \emph {et~al.}(1965)\citenamefont
  {Gielisse}, \citenamefont {Plendl}, \citenamefont {Mansur}, \citenamefont
  {Marshall}, \citenamefont {Mitra}, \citenamefont {Mykolajewycz},\ and\
  \citenamefont {Smakula}}]{JAP_borncharge_CoO}%
  \BibitemOpen
  \bibfield  {author} {\bibinfo {author} {\bibfnamefont {P.~J.}\ \bibnamefont
  {Gielisse}}, \bibinfo {author} {\bibfnamefont {J.~N.}\ \bibnamefont
  {Plendl}}, \bibinfo {author} {\bibfnamefont {L.~C.}\ \bibnamefont {Mansur}},
  \bibinfo {author} {\bibfnamefont {R.}~\bibnamefont {Marshall}}, \bibinfo
  {author} {\bibfnamefont {S.~S.}\ \bibnamefont {Mitra}}, \bibinfo {author}
  {\bibfnamefont {R.}~\bibnamefont {Mykolajewycz}}, \ and\ \bibinfo {author}
  {\bibfnamefont {A.}~\bibnamefont {Smakula}},\ }\href {\doibase
  10.1063/1.1714508} {\bibfield  {journal} {\bibinfo  {journal} {Journal of
  Applied Physics}\ }\textbf {\bibinfo {volume} {36}},\ \bibinfo {pages} {2446}
  (\bibinfo {year} {1965})}\BibitemShut {NoStop}%
\bibitem [{\citenamefont {Herrmann-Ronzaud}\ \emph {et~al.}(1978)\citenamefont
  {Herrmann-Ronzaud}, \citenamefont {Burlet},\ and\ \citenamefont
  {Rossat-Mignod}}]{CoO_magmoment}%
  \BibitemOpen
  \bibfield  {author} {\bibinfo {author} {\bibfnamefont {D.}~\bibnamefont
  {Herrmann-Ronzaud}}, \bibinfo {author} {\bibfnamefont {P.}~\bibnamefont
  {Burlet}}, \ and\ \bibinfo {author} {\bibfnamefont {J.}~\bibnamefont
  {Rossat-Mignod}},\ }\href {\doibase 10.1088/0022-3719/11/10/023} {\bibfield
  {journal} {\bibinfo  {journal} {Journal of Physics C: Solid State Physics}\
  }\textbf {\bibinfo {volume} {11}},\ \bibinfo {pages} {2123} (\bibinfo {year}
  {1978})}\BibitemShut {NoStop}%
\bibitem [{\citenamefont {Hüfner}\ \emph {et~al.}(1984)\citenamefont
  {Hüfner}, \citenamefont {Osterwalder}, \citenamefont {Riesterer},\ and\
  \citenamefont {Hulliger}}]{NiO_expt_gap}%
  \BibitemOpen
  \bibfield  {author} {\bibinfo {author} {\bibfnamefont {S.}~\bibnamefont
  {Hüfner}}, \bibinfo {author} {\bibfnamefont {J.}~\bibnamefont
  {Osterwalder}}, \bibinfo {author} {\bibfnamefont {T.}~\bibnamefont
  {Riesterer}}, \ and\ \bibinfo {author} {\bibfnamefont {F.}~\bibnamefont
  {Hulliger}},\ }\href {\doibase https://doi.org/10.1016/0038-1098(84)90007-3}
  {\bibfield  {journal} {\bibinfo  {journal} {Solid State Communications}\
  }\textbf {\bibinfo {volume} {52}},\ \bibinfo {pages} {793} (\bibinfo {year}
  {1984})}\BibitemShut {NoStop}%
\bibitem [{\citenamefont {Sawatzky}\ and\ \citenamefont
  {Allen}(1984)}]{NiO_expt_gap_prl1984}%
  \BibitemOpen
  \bibfield  {author} {\bibinfo {author} {\bibfnamefont {G.~A.}\ \bibnamefont
  {Sawatzky}}\ and\ \bibinfo {author} {\bibfnamefont {J.~W.}\ \bibnamefont
  {Allen}},\ }\href {\doibase 10.1103/PhysRevLett.53.2339} {\bibfield
  {journal} {\bibinfo  {journal} {Phys. Rev. Lett.}\ }\textbf {\bibinfo
  {volume} {53}},\ \bibinfo {pages} {2339} (\bibinfo {year}
  {1984})}\BibitemShut {NoStop}%
\bibitem [{\citenamefont {Wang}\ \emph {et~al.}(2010)\citenamefont {Wang},
  \citenamefont {Saal}, \citenamefont {Wang}, \citenamefont {Saengdeejing},
  \citenamefont {Shang}, \citenamefont {Chen},\ and\ \citenamefont
  {Liu}}]{Zhikun_Liiu_NiO}%
  \BibitemOpen
  \bibfield  {author} {\bibinfo {author} {\bibfnamefont {Y.}~\bibnamefont
  {Wang}}, \bibinfo {author} {\bibfnamefont {J.~E.}\ \bibnamefont {Saal}},
  \bibinfo {author} {\bibfnamefont {J.-J.}\ \bibnamefont {Wang}}, \bibinfo
  {author} {\bibfnamefont {A.}~\bibnamefont {Saengdeejing}}, \bibinfo {author}
  {\bibfnamefont {S.-L.}\ \bibnamefont {Shang}}, \bibinfo {author}
  {\bibfnamefont {L.-Q.}\ \bibnamefont {Chen}}, \ and\ \bibinfo {author}
  {\bibfnamefont {Z.-K.}\ \bibnamefont {Liu}},\ }\href {\doibase
  10.1103/PhysRevB.82.081104} {\bibfield  {journal} {\bibinfo  {journal} {Phys.
  Rev. B}\ }\textbf {\bibinfo {volume} {82}},\ \bibinfo {pages} {081104}
  (\bibinfo {year} {2010})}\BibitemShut {NoStop}%
\bibitem [{\citenamefont {Cheetham}\ and\ \citenamefont
  {Hope}(1983)}]{NiO_expt_magmoment}%
  \BibitemOpen
  \bibfield  {author} {\bibinfo {author} {\bibfnamefont {A.~K.}\ \bibnamefont
  {Cheetham}}\ and\ \bibinfo {author} {\bibfnamefont {D.~A.~O.}\ \bibnamefont
  {Hope}},\ }\href {\doibase 10.1103/PhysRevB.27.6964} {\bibfield  {journal}
  {\bibinfo  {journal} {Phys. Rev. B}\ }\textbf {\bibinfo {volume} {27}},\
  \bibinfo {pages} {6964} (\bibinfo {year} {1983})}\BibitemShut {NoStop}%
\bibitem [{\citenamefont {Fröhlich}(1954)}]{Frolichinteraction}%
  \BibitemOpen
  \bibfield  {author} {\bibinfo {author} {\bibfnamefont {H.}~\bibnamefont
  {Fröhlich}},\ }\href {\doibase 10.1080/00018735400101213} {\bibfield
  {journal} {\bibinfo  {journal} {Advances in Physics}\ }\textbf {\bibinfo
  {volume} {3}},\ \bibinfo {pages} {325} (\bibinfo {year} {1954})}\BibitemShut
  {NoStop}%
\bibitem [{\citenamefont {Zhou}\ \emph {et~al.}(2018)\citenamefont {Zhou},
  \citenamefont {Hellman},\ and\ \citenamefont {Bernardi}}]{STO_ZhouJinJian}%
  \BibitemOpen
  \bibfield  {author} {\bibinfo {author} {\bibfnamefont {J.-J.}\ \bibnamefont
  {Zhou}}, \bibinfo {author} {\bibfnamefont {O.}~\bibnamefont {Hellman}}, \
  and\ \bibinfo {author} {\bibfnamefont {M.}~\bibnamefont {Bernardi}},\ }\href
  {\doibase 10.1103/PhysRevLett.121.226603} {\bibfield  {journal} {\bibinfo
  {journal} {Phys. Rev. Lett.}\ }\textbf {\bibinfo {volume} {121}},\ \bibinfo
  {pages} {226603} (\bibinfo {year} {2018})}\BibitemShut {NoStop}%
\bibitem [{\citenamefont {Zhou}\ and\ \citenamefont
  {Bernardi}(2016)}]{GaAs_Zhou}%
  \BibitemOpen
  \bibfield  {author} {\bibinfo {author} {\bibfnamefont {J.-J.}\ \bibnamefont
  {Zhou}}\ and\ \bibinfo {author} {\bibfnamefont {M.}~\bibnamefont
  {Bernardi}},\ }\href {\doibase 10.1103/PhysRevB.94.201201} {\bibfield
  {journal} {\bibinfo  {journal} {Phys. Rev. B}\ }\textbf {\bibinfo {volume}
  {94}},\ \bibinfo {pages} {201201} (\bibinfo {year} {2016})}\BibitemShut
  {NoStop}%
\bibitem [{\citenamefont {Perdew}\ and\ \citenamefont
  {Zunger}(1981)}]{perdew1981self}%
  \BibitemOpen
  \bibfield  {author} {\bibinfo {author} {\bibfnamefont {J.~P.}\ \bibnamefont
  {Perdew}}\ and\ \bibinfo {author} {\bibfnamefont {A.}~\bibnamefont
  {Zunger}},\ }\href@noop {} {\bibfield  {journal} {\bibinfo  {journal}
  {Physical review B}\ }\textbf {\bibinfo {volume} {23}},\ \bibinfo {pages}
  {5048} (\bibinfo {year} {1981})}\BibitemShut {NoStop}%
\bibitem [{\citenamefont {Ghosez}\ \emph {et~al.}(1998)\citenamefont {Ghosez},
  \citenamefont {Michenaud},\ and\ \citenamefont {Gonze}}]{BornCharge_vasp1}%
  \BibitemOpen
  \bibfield  {author} {\bibinfo {author} {\bibfnamefont {P.}~\bibnamefont
  {Ghosez}}, \bibinfo {author} {\bibfnamefont {J.-P.}\ \bibnamefont
  {Michenaud}}, \ and\ \bibinfo {author} {\bibfnamefont {X.}~\bibnamefont
  {Gonze}},\ }\href {\doibase 10.1103/PhysRevB.58.6224} {\bibfield  {journal}
  {\bibinfo  {journal} {Phys. Rev. B}\ }\textbf {\bibinfo {volume} {58}},\
  \bibinfo {pages} {6224} (\bibinfo {year} {1998})}\BibitemShut {NoStop}%
\bibitem [{\citenamefont {Gonze}\ and\ \citenamefont
  {Lee}(1997)}]{BornCharge_vasp2}%
  \BibitemOpen
  \bibfield  {author} {\bibinfo {author} {\bibfnamefont {X.}~\bibnamefont
  {Gonze}}\ and\ \bibinfo {author} {\bibfnamefont {C.}~\bibnamefont {Lee}},\
  }\href {\doibase 10.1103/PhysRevB.55.10355} {\bibfield  {journal} {\bibinfo
  {journal} {Phys. Rev. B}\ }\textbf {\bibinfo {volume} {55}},\ \bibinfo
  {pages} {10355} (\bibinfo {year} {1997})}\BibitemShut {NoStop}%
\bibitem [{\citenamefont {King-Smith}\ and\ \citenamefont
  {Vanderbilt}(1993)}]{king1993theory}%
  \BibitemOpen
  \bibfield  {author} {\bibinfo {author} {\bibfnamefont {R.}~\bibnamefont
  {King-Smith}}\ and\ \bibinfo {author} {\bibfnamefont {D.}~\bibnamefont
  {Vanderbilt}},\ }\href {https://link.aps.org/doi/10.1103/PhysRevB.47.1651}
  {\bibfield  {journal} {\bibinfo  {journal} {Physical Review B}\ }\textbf
  {\bibinfo {volume} {47}},\ \bibinfo {pages} {1651} (\bibinfo {year}
  {1993})}\BibitemShut {NoStop}%
\bibitem [{\citenamefont {Resta}(1993)}]{resta1993macroscopic}%
  \BibitemOpen
  \bibfield  {author} {\bibinfo {author} {\bibfnamefont {R.}~\bibnamefont
  {Resta}},\ }\href@noop {} {\bibfield  {journal} {\bibinfo  {journal}
  {Europhysics Letters}\ }\textbf {\bibinfo {volume} {22}},\ \bibinfo {pages}
  {133} (\bibinfo {year} {1993})}\BibitemShut {NoStop}%
\bibitem [{\citenamefont {Onida}\ \emph {et~al.}(2002)\citenamefont {Onida},
  \citenamefont {Reining},\ and\ \citenamefont {Rubio}}]{Onida_RMP}%
  \BibitemOpen
  \bibfield  {author} {\bibinfo {author} {\bibfnamefont {G.}~\bibnamefont
  {Onida}}, \bibinfo {author} {\bibfnamefont {L.}~\bibnamefont {Reining}}, \
  and\ \bibinfo {author} {\bibfnamefont {A.}~\bibnamefont {Rubio}},\ }\href
  {\doibase 10.1103/RevModPhys.74.601} {\bibfield  {journal} {\bibinfo
  {journal} {Rev. Mod. Phys.}\ }\textbf {\bibinfo {volume} {74}},\ \bibinfo
  {pages} {601} (\bibinfo {year} {2002})}\BibitemShut {NoStop}%
\bibitem [{\citenamefont {Gajdo\ifmmode~\check{s}\else \v{s}\fi{}}\ \emph
  {et~al.}(2006)\citenamefont {Gajdo\ifmmode~\check{s}\else \v{s}\fi{}},
  \citenamefont {Hummer}, \citenamefont {Kresse}, \citenamefont
  {Furthm\"uller},\ and\ \citenamefont {Bechstedt}}]{Linear_optical_paw}%
  \BibitemOpen
  \bibfield  {author} {\bibinfo {author} {\bibfnamefont {M.}~\bibnamefont
  {Gajdo\ifmmode~\check{s}\else \v{s}\fi{}}}, \bibinfo {author} {\bibfnamefont
  {K.}~\bibnamefont {Hummer}}, \bibinfo {author} {\bibfnamefont
  {G.}~\bibnamefont {Kresse}}, \bibinfo {author} {\bibfnamefont
  {J.}~\bibnamefont {Furthm\"uller}}, \ and\ \bibinfo {author} {\bibfnamefont
  {F.}~\bibnamefont {Bechstedt}},\ }\href {\doibase 10.1103/PhysRevB.73.045112}
  {\bibfield  {journal} {\bibinfo  {journal} {Phys. Rev. B}\ }\textbf {\bibinfo
  {volume} {73}},\ \bibinfo {pages} {045112} (\bibinfo {year}
  {2006})}\BibitemShut {NoStop}%
\bibitem [{\citenamefont {Paier}\ \emph {et~al.}(2008)\citenamefont {Paier},
  \citenamefont {Marsman},\ and\ \citenamefont {Kresse}}]{Dielectric_HSE}%
  \BibitemOpen
  \bibfield  {author} {\bibinfo {author} {\bibfnamefont {J.}~\bibnamefont
  {Paier}}, \bibinfo {author} {\bibfnamefont {M.}~\bibnamefont {Marsman}}, \
  and\ \bibinfo {author} {\bibfnamefont {G.}~\bibnamefont {Kresse}},\ }\href
  {\doibase 10.1103/PhysRevB.78.121201} {\bibfield  {journal} {\bibinfo
  {journal} {Phys. Rev. B}\ }\textbf {\bibinfo {volume} {78}},\ \bibinfo
  {pages} {121201} (\bibinfo {year} {2008})}\BibitemShut {NoStop}%
\bibitem [{\citenamefont {Nazarov}\ and\ \citenamefont
  {Vignale}(2011)}]{Nazarov_exchangekernel}%
  \BibitemOpen
  \bibfield  {author} {\bibinfo {author} {\bibfnamefont {V.~U.}\ \bibnamefont
  {Nazarov}}\ and\ \bibinfo {author} {\bibfnamefont {G.}~\bibnamefont
  {Vignale}},\ }\href {\doibase 10.1103/PhysRevLett.107.216402} {\bibfield
  {journal} {\bibinfo  {journal} {Phys. Rev. Lett.}\ }\textbf {\bibinfo
  {volume} {107}},\ \bibinfo {pages} {216402} (\bibinfo {year}
  {2011})}\BibitemShut {NoStop}%
\bibitem [{\citenamefont {Penn}(1962)}]{penn1962wave}%
  \BibitemOpen
  \bibfield  {author} {\bibinfo {author} {\bibfnamefont {D.~R.}\ \bibnamefont
  {Penn}},\ }\href@noop {} {\bibfield  {journal} {\bibinfo  {journal} {Physical
  review}\ }\textbf {\bibinfo {volume} {128}},\ \bibinfo {pages} {2093}
  (\bibinfo {year} {1962})}\BibitemShut {NoStop}%
\bibitem [{\citenamefont {Onishi}\ and\ \citenamefont
  {Fu}(2024)}]{onishi2024universal}%
  \BibitemOpen
  \bibfield  {author} {\bibinfo {author} {\bibfnamefont {Y.}~\bibnamefont
  {Onishi}}\ and\ \bibinfo {author} {\bibfnamefont {L.}~\bibnamefont {Fu}},\
  }\href@noop {} {\bibfield  {journal} {\bibinfo  {journal} {Physical Review
  B}\ }\textbf {\bibinfo {volume} {110}},\ \bibinfo {pages} {155107} (\bibinfo
  {year} {2024})}\BibitemShut {NoStop}%
\bibitem [{\citenamefont {Perdew}\ and\ \citenamefont
  {Schmidt}(2001)}]{Jacob_ladder_DFT}%
  \BibitemOpen
  \bibfield  {author} {\bibinfo {author} {\bibfnamefont {J.~P.}\ \bibnamefont
  {Perdew}}\ and\ \bibinfo {author} {\bibfnamefont {K.}~\bibnamefont
  {Schmidt}},\ }\href {\doibase 10.1063/1.1390175} {\bibfield  {journal}
  {\bibinfo  {journal} {AIP Conference Proceedings}\ }\textbf {\bibinfo
  {volume} {577}},\ \bibinfo {pages} {1} (\bibinfo {year} {2001})}\BibitemShut
  {NoStop}%
\bibitem [{\citenamefont {Anisimov}\ \emph {et~al.}(1991)\citenamefont
  {Anisimov}, \citenamefont {Zaanen},\ and\ \citenamefont
  {Andersen}}]{LDAU_first}%
  \BibitemOpen
  \bibfield  {author} {\bibinfo {author} {\bibfnamefont {V.~I.}\ \bibnamefont
  {Anisimov}}, \bibinfo {author} {\bibfnamefont {J.}~\bibnamefont {Zaanen}}, \
  and\ \bibinfo {author} {\bibfnamefont {O.~K.}\ \bibnamefont {Andersen}},\
  }\href {\doibase 10.1103/PhysRevB.44.943} {\bibfield  {journal} {\bibinfo
  {journal} {Phys. Rev. B}\ }\textbf {\bibinfo {volume} {44}},\ \bibinfo
  {pages} {943} (\bibinfo {year} {1991})}\BibitemShut {NoStop}%
\bibitem [{\citenamefont {Anisimov}\ \emph {et~al.}(1993)\citenamefont
  {Anisimov}, \citenamefont {Solovyev}, \citenamefont {Korotin}, \citenamefont
  {Czy\ifmmode~\dot{z}\else \.{z}\fi{}yk},\ and\ \citenamefont
  {Sawatzky}}]{LDAU_Sawatzky}%
  \BibitemOpen
  \bibfield  {author} {\bibinfo {author} {\bibfnamefont {V.~I.}\ \bibnamefont
  {Anisimov}}, \bibinfo {author} {\bibfnamefont {I.~V.}\ \bibnamefont
  {Solovyev}}, \bibinfo {author} {\bibfnamefont {M.~A.}\ \bibnamefont
  {Korotin}}, \bibinfo {author} {\bibfnamefont {M.~T.}\ \bibnamefont
  {Czy\ifmmode~\dot{z}\else \.{z}\fi{}yk}}, \ and\ \bibinfo {author}
  {\bibfnamefont {G.~A.}\ \bibnamefont {Sawatzky}},\ }\href {\doibase
  10.1103/PhysRevB.48.16929} {\bibfield  {journal} {\bibinfo  {journal} {Phys.
  Rev. B}\ }\textbf {\bibinfo {volume} {48}},\ \bibinfo {pages} {16929}
  (\bibinfo {year} {1993})}\BibitemShut {NoStop}%
\bibitem [{\citenamefont {Heyd}\ \emph {et~al.}(2003)\citenamefont {Heyd},
  \citenamefont {Scuseria},\ and\ \citenamefont {Ernzerhof}}]{heyd2003hybrid}%
  \BibitemOpen
  \bibfield  {author} {\bibinfo {author} {\bibfnamefont {J.}~\bibnamefont
  {Heyd}}, \bibinfo {author} {\bibfnamefont {G.~E.}\ \bibnamefont {Scuseria}},
  \ and\ \bibinfo {author} {\bibfnamefont {M.}~\bibnamefont {Ernzerhof}},\
  }\href@noop {} {\bibfield  {journal} {\bibinfo  {journal} {The Journal of
  chemical physics}\ }\textbf {\bibinfo {volume} {118}},\ \bibinfo {pages}
  {8207} (\bibinfo {year} {2003})}\BibitemShut {NoStop}%
\bibitem [{\citenamefont {Nagamatsu}\ \emph {et~al.}(2001)\citenamefont
  {Nagamatsu}, \citenamefont {Nakagawa}, \citenamefont {Muranaka},
  \citenamefont {Zenitani},\ and\ \citenamefont
  {Akimitsu}}]{MgB2_Nagamatsu2001}%
  \BibitemOpen
  \bibfield  {author} {\bibinfo {author} {\bibfnamefont {J.}~\bibnamefont
  {Nagamatsu}}, \bibinfo {author} {\bibfnamefont {N.}~\bibnamefont {Nakagawa}},
  \bibinfo {author} {\bibfnamefont {T.}~\bibnamefont {Muranaka}}, \bibinfo
  {author} {\bibfnamefont {Y.}~\bibnamefont {Zenitani}}, \ and\ \bibinfo
  {author} {\bibfnamefont {J.}~\bibnamefont {Akimitsu}},\ }\href {\doibase
  10.1038/35065039} {\bibfield  {journal} {\bibinfo  {journal} {Nature}\
  }\textbf {\bibinfo {volume} {410}},\ \bibinfo {pages} {63} (\bibinfo {year}
  {2001})}\BibitemShut {NoStop}%
\bibitem [{\citenamefont {Chen}\ \emph {et~al.}(2001)\citenamefont {Chen},
  \citenamefont {Konstantinovi\ifmmode~\acute{c}\else \'{c}\fi{}},
  \citenamefont {Irwin}, \citenamefont {Lawrie},\ and\ \citenamefont
  {Franck}}]{Twogap_MgB2}%
  \BibitemOpen
  \bibfield  {author} {\bibinfo {author} {\bibfnamefont {X.~K.}\ \bibnamefont
  {Chen}}, \bibinfo {author} {\bibfnamefont {M.~J.}\ \bibnamefont
  {Konstantinovi\ifmmode~\acute{c}\else \'{c}\fi{}}}, \bibinfo {author}
  {\bibfnamefont {J.~C.}\ \bibnamefont {Irwin}}, \bibinfo {author}
  {\bibfnamefont {D.~D.}\ \bibnamefont {Lawrie}}, \ and\ \bibinfo {author}
  {\bibfnamefont {J.~P.}\ \bibnamefont {Franck}},\ }\href {\doibase
  10.1103/PhysRevLett.87.157002} {\bibfield  {journal} {\bibinfo  {journal}
  {Phys. Rev. Lett.}\ }\textbf {\bibinfo {volume} {87}},\ \bibinfo {pages}
  {157002} (\bibinfo {year} {2001})}\BibitemShut {NoStop}%
\bibitem [{\citenamefont {Tsuda}\ \emph {et~al.}(2001)\citenamefont {Tsuda},
  \citenamefont {Yokoya}, \citenamefont {Kiss}, \citenamefont {Takano},
  \citenamefont {Togano}, \citenamefont {Kito}, \citenamefont {Ihara},\ and\
  \citenamefont {Shin}}]{MgB2_MGPRL}%
  \BibitemOpen
  \bibfield  {author} {\bibinfo {author} {\bibfnamefont {S.}~\bibnamefont
  {Tsuda}}, \bibinfo {author} {\bibfnamefont {T.}~\bibnamefont {Yokoya}},
  \bibinfo {author} {\bibfnamefont {T.}~\bibnamefont {Kiss}}, \bibinfo {author}
  {\bibfnamefont {Y.}~\bibnamefont {Takano}}, \bibinfo {author} {\bibfnamefont
  {K.}~\bibnamefont {Togano}}, \bibinfo {author} {\bibfnamefont
  {H.}~\bibnamefont {Kito}}, \bibinfo {author} {\bibfnamefont {H.}~\bibnamefont
  {Ihara}}, \ and\ \bibinfo {author} {\bibfnamefont {S.}~\bibnamefont {Shin}},\
  }\href {\doibase 10.1103/PhysRevLett.87.177006} {\bibfield  {journal}
  {\bibinfo  {journal} {Phys. Rev. Lett.}\ }\textbf {\bibinfo {volume} {87}},\
  \bibinfo {pages} {177006} (\bibinfo {year} {2001})}\BibitemShut {NoStop}%
\bibitem [{\citenamefont {Baron}\ \emph {et~al.}(2004)\citenamefont {Baron},
  \citenamefont {Uchiyama}, \citenamefont {Tanaka}, \citenamefont {Tsutsui},
  \citenamefont {Ishikawa}, \citenamefont {Lee}, \citenamefont {Heid},
  \citenamefont {Bohnen}, \citenamefont {Tajima},\ and\ \citenamefont
  {Ishikawa}}]{Kogn_anomaly_MgB2}%
  \BibitemOpen
  \bibfield  {author} {\bibinfo {author} {\bibfnamefont {A.~Q.~R.}\
  \bibnamefont {Baron}}, \bibinfo {author} {\bibfnamefont {H.}~\bibnamefont
  {Uchiyama}}, \bibinfo {author} {\bibfnamefont {Y.}~\bibnamefont {Tanaka}},
  \bibinfo {author} {\bibfnamefont {S.}~\bibnamefont {Tsutsui}}, \bibinfo
  {author} {\bibfnamefont {D.}~\bibnamefont {Ishikawa}}, \bibinfo {author}
  {\bibfnamefont {S.}~\bibnamefont {Lee}}, \bibinfo {author} {\bibfnamefont
  {R.}~\bibnamefont {Heid}}, \bibinfo {author} {\bibfnamefont {K.-P.}\
  \bibnamefont {Bohnen}}, \bibinfo {author} {\bibfnamefont {S.}~\bibnamefont
  {Tajima}}, \ and\ \bibinfo {author} {\bibfnamefont {T.}~\bibnamefont
  {Ishikawa}},\ }\href {\doibase 10.1103/PhysRevLett.92.197004} {\bibfield
  {journal} {\bibinfo  {journal} {Phys. Rev. Lett.}\ }\textbf {\bibinfo
  {volume} {92}},\ \bibinfo {pages} {197004} (\bibinfo {year}
  {2004})}\BibitemShut {NoStop}%
\bibitem [{\citenamefont {d'Astuto}\ \emph
  {et~al.}(2007{\natexlab{a}})\citenamefont {d'Astuto}, \citenamefont
  {Calandra}, \citenamefont {Reich}, \citenamefont {Shukla}, \citenamefont
  {Lazzeri}, \citenamefont {Mauri}, \citenamefont {Karpinski}, \citenamefont
  {Zhigadlo}, \citenamefont {Bossak},\ and\ \citenamefont
  {Krisch}}]{WeakanharmonicMgB2}%
  \BibitemOpen
  \bibfield  {author} {\bibinfo {author} {\bibfnamefont {M.}~\bibnamefont
  {d'Astuto}}, \bibinfo {author} {\bibfnamefont {M.}~\bibnamefont {Calandra}},
  \bibinfo {author} {\bibfnamefont {S.}~\bibnamefont {Reich}}, \bibinfo
  {author} {\bibfnamefont {A.}~\bibnamefont {Shukla}}, \bibinfo {author}
  {\bibfnamefont {M.}~\bibnamefont {Lazzeri}}, \bibinfo {author} {\bibfnamefont
  {F.}~\bibnamefont {Mauri}}, \bibinfo {author} {\bibfnamefont
  {J.}~\bibnamefont {Karpinski}}, \bibinfo {author} {\bibfnamefont {N.~D.}\
  \bibnamefont {Zhigadlo}}, \bibinfo {author} {\bibfnamefont {A.}~\bibnamefont
  {Bossak}}, \ and\ \bibinfo {author} {\bibfnamefont {M.}~\bibnamefont
  {Krisch}},\ }\href {\doibase 10.1103/PhysRevB.75.174508} {\bibfield
  {journal} {\bibinfo  {journal} {Phys. Rev. B}\ }\textbf {\bibinfo {volume}
  {75}},\ \bibinfo {pages} {174508} (\bibinfo {year}
  {2007}{\natexlab{a}})}\BibitemShut {NoStop}%
\bibitem [{\citenamefont {Choi}\ \emph
  {et~al.}(2002{\natexlab{a}})\citenamefont {Choi}, \citenamefont {Roundy},
  \citenamefont {Sun}, \citenamefont {Cohen},\ and\ \citenamefont
  {Louie}}]{LouieMgB2nature}%
  \BibitemOpen
  \bibfield  {author} {\bibinfo {author} {\bibfnamefont {H.~J.}\ \bibnamefont
  {Choi}}, \bibinfo {author} {\bibfnamefont {D.}~\bibnamefont {Roundy}},
  \bibinfo {author} {\bibfnamefont {H.}~\bibnamefont {Sun}}, \bibinfo {author}
  {\bibfnamefont {M.~L.}\ \bibnamefont {Cohen}}, \ and\ \bibinfo {author}
  {\bibfnamefont {S.~G.}\ \bibnamefont {Louie}},\ }\href {\doibase
  10.1038/nature00898} {\bibfield  {journal} {\bibinfo  {journal} {Nature}\
  }\textbf {\bibinfo {volume} {418}},\ \bibinfo {pages} {758} (\bibinfo {year}
  {2002}{\natexlab{a}})}\BibitemShut {NoStop}%
\bibitem [{\citenamefont {Margine}\ and\ \citenamefont
  {Giustino}(2013)}]{Anisotropic_ME}%
  \BibitemOpen
  \bibfield  {author} {\bibinfo {author} {\bibfnamefont {E.~R.}\ \bibnamefont
  {Margine}}\ and\ \bibinfo {author} {\bibfnamefont {F.}~\bibnamefont
  {Giustino}},\ }\href {\doibase 10.1103/PhysRevB.87.024505} {\bibfield
  {journal} {\bibinfo  {journal} {Phys. Rev. B}\ }\textbf {\bibinfo {volume}
  {87}},\ \bibinfo {pages} {024505} (\bibinfo {year} {2013})}\BibitemShut
  {NoStop}%
\bibitem [{\citenamefont {Kortus}\ \emph {et~al.}(2001)\citenamefont {Kortus},
  \citenamefont {Mazin}, \citenamefont {Belashchenko}, \citenamefont
  {Antropov},\ and\ \citenamefont {Boyer}}]{Kortus_MgB2}%
  \BibitemOpen
  \bibfield  {author} {\bibinfo {author} {\bibfnamefont {J.}~\bibnamefont
  {Kortus}}, \bibinfo {author} {\bibfnamefont {I.~I.}\ \bibnamefont {Mazin}},
  \bibinfo {author} {\bibfnamefont {K.~D.}\ \bibnamefont {Belashchenko}},
  \bibinfo {author} {\bibfnamefont {V.~P.}\ \bibnamefont {Antropov}}, \ and\
  \bibinfo {author} {\bibfnamefont {L.~L.}\ \bibnamefont {Boyer}},\ }\href
  {\doibase 10.1103/PhysRevLett.86.4656} {\bibfield  {journal} {\bibinfo
  {journal} {Phys. Rev. Lett.}\ }\textbf {\bibinfo {volume} {86}},\ \bibinfo
  {pages} {4656} (\bibinfo {year} {2001})}\BibitemShut {NoStop}%
\bibitem [{\citenamefont {Yildirim}\ \emph {et~al.}(2001)\citenamefont
  {Yildirim}, \citenamefont {G\"ulseren}, \citenamefont {Lynn}, \citenamefont
  {Brown}, \citenamefont {Udovic}, \citenamefont {Huang}, \citenamefont
  {Rogado}, \citenamefont {Regan}, \citenamefont {Hayward}, \citenamefont
  {Slusky}, \citenamefont {He}, \citenamefont {Haas}, \citenamefont {Khalifah},
  \citenamefont {Inumaru},\ and\ \citenamefont {Cava}}]{Giant_Anharmon_MgB2}%
  \BibitemOpen
  \bibfield  {author} {\bibinfo {author} {\bibfnamefont {T.}~\bibnamefont
  {Yildirim}}, \bibinfo {author} {\bibfnamefont {O.}~\bibnamefont
  {G\"ulseren}}, \bibinfo {author} {\bibfnamefont {J.~W.}\ \bibnamefont
  {Lynn}}, \bibinfo {author} {\bibfnamefont {C.~M.}\ \bibnamefont {Brown}},
  \bibinfo {author} {\bibfnamefont {T.~J.}\ \bibnamefont {Udovic}}, \bibinfo
  {author} {\bibfnamefont {Q.}~\bibnamefont {Huang}}, \bibinfo {author}
  {\bibfnamefont {N.}~\bibnamefont {Rogado}}, \bibinfo {author} {\bibfnamefont
  {K.~A.}\ \bibnamefont {Regan}}, \bibinfo {author} {\bibfnamefont {M.~A.}\
  \bibnamefont {Hayward}}, \bibinfo {author} {\bibfnamefont {J.~S.}\
  \bibnamefont {Slusky}}, \bibinfo {author} {\bibfnamefont {T.}~\bibnamefont
  {He}}, \bibinfo {author} {\bibfnamefont {M.~K.}\ \bibnamefont {Haas}},
  \bibinfo {author} {\bibfnamefont {P.}~\bibnamefont {Khalifah}}, \bibinfo
  {author} {\bibfnamefont {K.}~\bibnamefont {Inumaru}}, \ and\ \bibinfo
  {author} {\bibfnamefont {R.~J.}\ \bibnamefont {Cava}},\ }\href {\doibase
  10.1103/PhysRevLett.87.037001} {\bibfield  {journal} {\bibinfo  {journal}
  {Phys. Rev. Lett.}\ }\textbf {\bibinfo {volume} {87}},\ \bibinfo {pages}
  {037001} (\bibinfo {year} {2001})}\BibitemShut {NoStop}%
\bibitem [{\citenamefont {Choi}\ \emph
  {et~al.}(2002{\natexlab{b}})\citenamefont {Choi}, \citenamefont {Roundy},
  \citenamefont {Sun}, \citenamefont {Cohen},\ and\ \citenamefont
  {Louie}}]{Choi_PRB_Louie}%
  \BibitemOpen
  \bibfield  {author} {\bibinfo {author} {\bibfnamefont {H.~J.}\ \bibnamefont
  {Choi}}, \bibinfo {author} {\bibfnamefont {D.}~\bibnamefont {Roundy}},
  \bibinfo {author} {\bibfnamefont {H.}~\bibnamefont {Sun}}, \bibinfo {author}
  {\bibfnamefont {M.~L.}\ \bibnamefont {Cohen}}, \ and\ \bibinfo {author}
  {\bibfnamefont {S.~G.}\ \bibnamefont {Louie}},\ }\href {\doibase
  10.1103/PhysRevB.66.020513} {\bibfield  {journal} {\bibinfo  {journal} {Phys.
  Rev. B}\ }\textbf {\bibinfo {volume} {66}},\ \bibinfo {pages} {020513}
  (\bibinfo {year} {2002}{\natexlab{b}})}\BibitemShut {NoStop}%
\bibitem [{\citenamefont {Guritanu}\ \emph {et~al.}(2006)\citenamefont
  {Guritanu}, \citenamefont {Kuzmenko}, \citenamefont {van~der Marel},
  \citenamefont {Kazakov}, \citenamefont {Zhigadlo},\ and\ \citenamefont
  {Karpinski}}]{Anisotropic_optical_MgB2}%
  \BibitemOpen
  \bibfield  {author} {\bibinfo {author} {\bibfnamefont {V.}~\bibnamefont
  {Guritanu}}, \bibinfo {author} {\bibfnamefont {A.~B.}\ \bibnamefont
  {Kuzmenko}}, \bibinfo {author} {\bibfnamefont {D.}~\bibnamefont {van~der
  Marel}}, \bibinfo {author} {\bibfnamefont {S.~M.}\ \bibnamefont {Kazakov}},
  \bibinfo {author} {\bibfnamefont {N.~D.}\ \bibnamefont {Zhigadlo}}, \ and\
  \bibinfo {author} {\bibfnamefont {J.}~\bibnamefont {Karpinski}},\ }\href
  {\doibase 10.1103/PhysRevB.73.104509} {\bibfield  {journal} {\bibinfo
  {journal} {Phys. Rev. B}\ }\textbf {\bibinfo {volume} {73}},\ \bibinfo
  {pages} {104509} (\bibinfo {year} {2006})}\BibitemShut {NoStop}%
\bibitem [{\citenamefont {Hlinka}\ \emph {et~al.}(2001)\citenamefont {Hlinka},
  \citenamefont {Gregora}, \citenamefont {Pokorn\'y}, \citenamefont {Plecenik},
  \citenamefont {K\'u\ifmmode~\check{s}\else \v{s}\fi{}}, \citenamefont
  {Satrapinsky},\ and\ \citenamefont {Be\ifmmode \check{n}\else
  \v{n}\fi{}a\ifmmode~\check{c}\else \v{c}\fi{}ka}}]{Raman_mgB2_prb}%
  \BibitemOpen
  \bibfield  {author} {\bibinfo {author} {\bibfnamefont {J.}~\bibnamefont
  {Hlinka}}, \bibinfo {author} {\bibfnamefont {I.}~\bibnamefont {Gregora}},
  \bibinfo {author} {\bibfnamefont {J.}~\bibnamefont {Pokorn\'y}}, \bibinfo
  {author} {\bibfnamefont {A.}~\bibnamefont {Plecenik}}, \bibinfo {author}
  {\bibfnamefont {P.}~\bibnamefont {K\'u\ifmmode~\check{s}\else \v{s}\fi{}}},
  \bibinfo {author} {\bibfnamefont {L.}~\bibnamefont {Satrapinsky}}, \ and\
  \bibinfo {author} {\bibfnamefont {i.~c.~v.}\ \bibnamefont {Be\ifmmode
  \check{n}\else \v{n}\fi{}a\ifmmode~\check{c}\else \v{c}\fi{}ka}},\ }\href
  {\doibase 10.1103/PhysRevB.64.140503} {\bibfield  {journal} {\bibinfo
  {journal} {Phys. Rev. B}\ }\textbf {\bibinfo {volume} {64}},\ \bibinfo
  {pages} {140503} (\bibinfo {year} {2001})}\BibitemShut {NoStop}%
\bibitem [{\citenamefont {Goncharov}\ \emph {et~al.}(2001)\citenamefont
  {Goncharov}, \citenamefont {Struzhkin}, \citenamefont {Gregoryanz},
  \citenamefont {Hu}, \citenamefont {Hemley}, \citenamefont {Mao},
  \citenamefont {Lapertot}, \citenamefont {Bud'ko},\ and\ \citenamefont
  {Canfield}}]{Raman_mgB2_pressure}%
  \BibitemOpen
  \bibfield  {author} {\bibinfo {author} {\bibfnamefont {A.~F.}\ \bibnamefont
  {Goncharov}}, \bibinfo {author} {\bibfnamefont {V.~V.}\ \bibnamefont
  {Struzhkin}}, \bibinfo {author} {\bibfnamefont {E.}~\bibnamefont
  {Gregoryanz}}, \bibinfo {author} {\bibfnamefont {J.}~\bibnamefont {Hu}},
  \bibinfo {author} {\bibfnamefont {R.~J.}\ \bibnamefont {Hemley}}, \bibinfo
  {author} {\bibfnamefont {H.-k.}\ \bibnamefont {Mao}}, \bibinfo {author}
  {\bibfnamefont {G.}~\bibnamefont {Lapertot}}, \bibinfo {author}
  {\bibfnamefont {S.~L.}\ \bibnamefont {Bud'ko}}, \ and\ \bibinfo {author}
  {\bibfnamefont {P.~C.}\ \bibnamefont {Canfield}},\ }\href {\doibase
  10.1103/PhysRevB.64.100509} {\bibfield  {journal} {\bibinfo  {journal} {Phys.
  Rev. B}\ }\textbf {\bibinfo {volume} {64}},\ \bibinfo {pages} {100509}
  (\bibinfo {year} {2001})}\BibitemShut {NoStop}%
\bibitem [{\citenamefont {d'Astuto}\ \emph
  {et~al.}(2007{\natexlab{b}})\citenamefont {d'Astuto}, \citenamefont
  {Calandra}, \citenamefont {Reich}, \citenamefont {Shukla}, \citenamefont
  {Lazzeri}, \citenamefont {Mauri}, \citenamefont {Karpinski}, \citenamefont
  {Zhigadlo}, \citenamefont {Bossak},\ and\ \citenamefont
  {Krisch}}]{MgB2_phexpt}%
  \BibitemOpen
  \bibfield  {author} {\bibinfo {author} {\bibfnamefont {M.}~\bibnamefont
  {d'Astuto}}, \bibinfo {author} {\bibfnamefont {M.}~\bibnamefont {Calandra}},
  \bibinfo {author} {\bibfnamefont {S.}~\bibnamefont {Reich}}, \bibinfo
  {author} {\bibfnamefont {A.}~\bibnamefont {Shukla}}, \bibinfo {author}
  {\bibfnamefont {M.}~\bibnamefont {Lazzeri}}, \bibinfo {author} {\bibfnamefont
  {F.}~\bibnamefont {Mauri}}, \bibinfo {author} {\bibfnamefont
  {J.}~\bibnamefont {Karpinski}}, \bibinfo {author} {\bibfnamefont {N.~D.}\
  \bibnamefont {Zhigadlo}}, \bibinfo {author} {\bibfnamefont {A.}~\bibnamefont
  {Bossak}}, \ and\ \bibinfo {author} {\bibfnamefont {M.}~\bibnamefont
  {Krisch}},\ }\href {\doibase 10.1103/PhysRevB.75.174508} {\bibfield
  {journal} {\bibinfo  {journal} {Phys. Rev. B}\ }\textbf {\bibinfo {volume}
  {75}},\ \bibinfo {pages} {174508} (\bibinfo {year}
  {2007}{\natexlab{b}})}\BibitemShut {NoStop}%
\bibitem [{\citenamefont {Floris}\ \emph {et~al.}(2005)\citenamefont {Floris},
  \citenamefont {Profeta}, \citenamefont {Lathiotakis}, \citenamefont
  {L\"uders}, \citenamefont {Marques}, \citenamefont {Franchini}, \citenamefont
  {Gross}, \citenamefont {Continenza},\ and\ \citenamefont
  {Massidda}}]{FlorisSCDFT}%
  \BibitemOpen
  \bibfield  {author} {\bibinfo {author} {\bibfnamefont {A.}~\bibnamefont
  {Floris}}, \bibinfo {author} {\bibfnamefont {G.}~\bibnamefont {Profeta}},
  \bibinfo {author} {\bibfnamefont {N.~N.}\ \bibnamefont {Lathiotakis}},
  \bibinfo {author} {\bibfnamefont {M.}~\bibnamefont {L\"uders}}, \bibinfo
  {author} {\bibfnamefont {M.~A.~L.}\ \bibnamefont {Marques}}, \bibinfo
  {author} {\bibfnamefont {C.}~\bibnamefont {Franchini}}, \bibinfo {author}
  {\bibfnamefont {E.~K.~U.}\ \bibnamefont {Gross}}, \bibinfo {author}
  {\bibfnamefont {A.}~\bibnamefont {Continenza}}, \ and\ \bibinfo {author}
  {\bibfnamefont {S.}~\bibnamefont {Massidda}},\ }\href {\doibase
  10.1103/PhysRevLett.94.037004} {\bibfield  {journal} {\bibinfo  {journal}
  {Phys. Rev. Lett.}\ }\textbf {\bibinfo {volume} {94}},\ \bibinfo {pages}
  {037004} (\bibinfo {year} {2005})}\BibitemShut {NoStop}%
\bibitem [{\citenamefont {Kresse}\ and\ \citenamefont
  {Joubert}(1999)}]{Kresse1999}%
  \BibitemOpen
  \bibfield  {author} {\bibinfo {author} {\bibfnamefont {G.}~\bibnamefont
  {Kresse}}\ and\ \bibinfo {author} {\bibfnamefont {D.}~\bibnamefont
  {Joubert}},\ }\href {\doibase 10.1103/PhysRevB.59.1758} {\bibfield  {journal}
  {\bibinfo  {journal} {Physical Review B}\ }\textbf {\bibinfo {volume} {59}},\
  \bibinfo {pages} {1758} (\bibinfo {year} {1999})}\BibitemShut {NoStop}%
\bibitem [{\citenamefont {Kresse}\ and\ \citenamefont
  {Hafner}(1993)}]{Kresse1993}%
  \BibitemOpen
  \bibfield  {author} {\bibinfo {author} {\bibfnamefont {G.}~\bibnamefont
  {Kresse}}\ and\ \bibinfo {author} {\bibfnamefont {J.}~\bibnamefont
  {Hafner}},\ }\href {\doibase 10.1103/PhysRevB.48.13115} {\bibfield  {journal}
  {\bibinfo  {journal} {Physical Review B}\ }\textbf {\bibinfo {volume} {48}},\
  \bibinfo {pages} {13115} (\bibinfo {year} {1993})}\BibitemShut {NoStop}%
\bibitem [{\citenamefont {Kresse}\ and\ \citenamefont
  {Furthm{\"{u}}ller}(1996)}]{Kresse1996}%
  \BibitemOpen
  \bibfield  {author} {\bibinfo {author} {\bibfnamefont {G.}~\bibnamefont
  {Kresse}}\ and\ \bibinfo {author} {\bibfnamefont {J.}~\bibnamefont
  {Furthm{\"{u}}ller}},\ }\href {\doibase 10.1103/PhysRevB.54.11169} {\bibfield
   {journal} {\bibinfo  {journal} {Physical Review B}\ }\textbf {\bibinfo
  {volume} {54}},\ \bibinfo {pages} {11169} (\bibinfo {year}
  {1996})}\BibitemShut {NoStop}%
\bibitem [{\citenamefont {Chaput}\ \emph {et~al.}(2019)\citenamefont {Chaput},
  \citenamefont {Togo},\ and\ \citenamefont {Tanaka}}]{FDEPCPAW}%
  \BibitemOpen
  \bibfield  {author} {\bibinfo {author} {\bibfnamefont {L.}~\bibnamefont
  {Chaput}}, \bibinfo {author} {\bibfnamefont {A.}~\bibnamefont {Togo}}, \ and\
  \bibinfo {author} {\bibfnamefont {I.}~\bibnamefont {Tanaka}},\ }\href
  {\doibase 10.1103/PhysRevB.100.174304} {\bibfield  {journal} {\bibinfo
  {journal} {Phys. Rev. B}\ }\textbf {\bibinfo {volume} {100}},\ \bibinfo
  {pages} {174304} (\bibinfo {year} {2019})}\BibitemShut {NoStop}%
\bibitem [{\citenamefont {Mostofi}\ \emph {et~al.}(2014)\citenamefont
  {Mostofi}, \citenamefont {Yates}, \citenamefont {Pizzi}, \citenamefont {Lee},
  \citenamefont {Souza}, \citenamefont {Vanderbilt},\ and\ \citenamefont
  {Marzari}}]{MOSTOFI20142309}%
  \BibitemOpen
  \bibfield  {author} {\bibinfo {author} {\bibfnamefont {A.~A.}\ \bibnamefont
  {Mostofi}}, \bibinfo {author} {\bibfnamefont {J.~R.}\ \bibnamefont {Yates}},
  \bibinfo {author} {\bibfnamefont {G.}~\bibnamefont {Pizzi}}, \bibinfo
  {author} {\bibfnamefont {Y.-S.}\ \bibnamefont {Lee}}, \bibinfo {author}
  {\bibfnamefont {I.}~\bibnamefont {Souza}}, \bibinfo {author} {\bibfnamefont
  {D.}~\bibnamefont {Vanderbilt}}, \ and\ \bibinfo {author} {\bibfnamefont
  {N.}~\bibnamefont {Marzari}},\ }\href {\doibase
  https://doi.org/10.1016/j.cpc.2014.05.003} {\bibfield  {journal} {\bibinfo
  {journal} {Computer Physics Communications}\ }\textbf {\bibinfo {volume}
  {185}},\ \bibinfo {pages} {2309} (\bibinfo {year} {2014})}\BibitemShut
  {NoStop}%
\bibitem [{\citenamefont {Migdal}(1958)}]{Migdal1958}%
  \BibitemOpen
  \bibfield  {author} {\bibinfo {author} {\bibfnamefont {A.~B.}\ \bibnamefont
  {Migdal}},\ }\href {http://jetp.ras.ru/cgi-bin/e/index/e/7/6/p996?a=list}
  {\bibfield  {journal} {\bibinfo  {journal} {Sov. Phys. JETP}\ }\textbf
  {\bibinfo {volume} {7}},\ \bibinfo {pages} {996} (\bibinfo {year}
  {1958})}\BibitemShut {NoStop}%
\bibitem [{\citenamefont {Marini}\ \emph {et~al.}(2024)\citenamefont {Marini},
  \citenamefont {Marchese}, \citenamefont {Profeta}, \citenamefont {Sjakste},
  \citenamefont {Macheda}, \citenamefont {Vast}, \citenamefont {Mauri},\ and\
  \citenamefont {Calandra}}]{epiq_impl2024}%
  \BibitemOpen
  \bibfield  {author} {\bibinfo {author} {\bibfnamefont {G.}~\bibnamefont
  {Marini}}, \bibinfo {author} {\bibfnamefont {G.}~\bibnamefont {Marchese}},
  \bibinfo {author} {\bibfnamefont {G.}~\bibnamefont {Profeta}}, \bibinfo
  {author} {\bibfnamefont {J.}~\bibnamefont {Sjakste}}, \bibinfo {author}
  {\bibfnamefont {F.}~\bibnamefont {Macheda}}, \bibinfo {author} {\bibfnamefont
  {N.}~\bibnamefont {Vast}}, \bibinfo {author} {\bibfnamefont {F.}~\bibnamefont
  {Mauri}}, \ and\ \bibinfo {author} {\bibfnamefont {M.}~\bibnamefont
  {Calandra}},\ }\href {\doibase https://doi.org/10.1016/j.cpc.2023.108950}
  {\bibfield  {journal} {\bibinfo  {journal} {Computer Physics Communications}\
  }\textbf {\bibinfo {volume} {295}},\ \bibinfo {pages} {108950} (\bibinfo
  {year} {2024})}\BibitemShut {NoStop}%
\bibitem [{\citenamefont {McMillan}(1968)}]{McMillan1968}%
  \BibitemOpen
  \bibfield  {author} {\bibinfo {author} {\bibfnamefont {W.~L.}\ \bibnamefont
  {McMillan}},\ }\href {\doibase 10.1103/PhysRev.167.331} {\bibfield  {journal}
  {\bibinfo  {journal} {Phys. Rev.}\ }\textbf {\bibinfo {volume} {167}},\
  \bibinfo {pages} {331} (\bibinfo {year} {1968})}\BibitemShut {NoStop}%
\bibitem [{\citenamefont {Allen}\ and\ \citenamefont
  {Dynes}(1975)}]{Allen1975}%
  \BibitemOpen
  \bibfield  {author} {\bibinfo {author} {\bibfnamefont {P.~B.}\ \bibnamefont
  {Allen}}\ and\ \bibinfo {author} {\bibfnamefont {R.~C.}\ \bibnamefont
  {Dynes}},\ }\href {\doibase 10.1103/PhysRevB.12.905} {\bibfield  {journal}
  {\bibinfo  {journal} {Phys. Rev. B}\ }\textbf {\bibinfo {volume} {12}},\
  \bibinfo {pages} {905} (\bibinfo {year} {1975})}\BibitemShut {NoStop}%
\end{thebibliography}%

\SB{Acknowledgements}\textbf{\---}Y.W., R.Z., and J.S. acknowledge the support of the U.S. Office of Naval Research (ONR) Grant No. N00014-22-1-2673.  The work at Tulane University was also supported by the Advanced Cyberinfrastructure Coordination Ecosystem funded by the National Science Foundation, and the National Energy Research Scientific Computing Center (NERSC) using NERSC Awards No. BES-ERCAP0020494 and No. BESERCAP0028709. The work at Northeastern University was supported by the National Science Foundation through the Expand-QISE award NSF-OMA-2329067 and benefited from the resources of Northeastern University’s Advanced Scientific Computation Center, the Discovery Cluster, and the Massachusetts Technology Collaborative award MTC-22032. The work at Los Alamos National Laboratory was carried out under the auspices of the U.S. Department of Energy (DOE) National Nuclear Security Administration under Contract No. 89233218CNA000001. It was supported by the LANL LDRD Program, the Quantum Science Center, a U.S. DOE Office of Science National Quantum Information Science Research Center, and in part by the Center for Integrated Nanotechnologies, a DOE BES user facility, in partnership with the LANL Institutional Computing Program for computational resources. Additional computations were performed at the NERSC, a U.S. Department of Energy Office of Science User
Facility located at Lawrence Berkeley National Laboratory, operated under Contract No. DE-AC02-
05CH11231 using NERSC Award No. ERCAP0020494. B. B. acknowledges support from the COST Action
CA16218.

%\section*{Author contributions}
\SB{Author contributions}\textbf{\---}R.Z., J.S., and A.B. designed the study. Y.W, M.E., and R.Z. performed first principles computations and analyzed the data with help from C.L. H.M., B.B,  R.S.M., J.Z, G. K., A.B., and J.S. A.B., and J.S. provided computational infrastructure. R.Z., Y.W, C.L., B.B, R.S.M., A.B. and J.S. wrote the manuscript with input from all the authors. All authors contributed to editing the manuscript.
%\section*{Additional information}

\SB{Additional information}\textbf{\---}The authors declare no competing financial interests. 

\end{document}